\documentclass[aps, prfluids, preprint, superscriptaddress, longbibliography]{revtex4-1}

\setcitestyle{numbers,square,articletitle=true}
\usepackage{epsfig}
\usepackage[normalem]{ulem}
\usepackage{graphicx}
\usepackage{amssymb}
\usepackage{amsmath}
\usepackage{comment}
\usepackage{textcomp} 
\usepackage{color, soul}
\usepackage{mathtools}
\usepackage{gensymb}
\usepackage{subcaption}
\usepackage{epstopdf}
\usepackage{adjustbox}
\usepackage{booktabs}
\usepackage{array}
\usepackage{multirow}
\usepackage{float}
\usepackage{tabularx}
\definecolor{FuchsiaPink}{rgb}{0.78, 0.08, 0.52}
\usepackage[T1]{fontenc}
\usepackage{lmodern}
\usepackage{microtype}

\usepackage[capitalize]{cleveref}

\begin{document}

\title{Penetration of a sphere in a pool: Secondary pinch-off}
    \author{Prasanna Kumar Billa}
\affiliation{Multiscale Multiphysics Group (MMG), Department of Mechanical Engineering, Indian Institute of Technology Madras, 600036, India}

\author{Cameron Tropea}
 \email{tropea@sla.tu-darmstadt.de}
 \affiliation{Department of Mechanical Engineering, Indian Institute of Technology Madras, 600036, India}
\affiliation{Institute for Fluid Mechanics and Aerodynamics, Technical University of Darmstadt, 64287, Germany}

 \author{Pallab Sinha Mahapatra}
\email{pallab@iitm.ac.in}
\affiliation{Multiscale Multiphysics Group (MMG), Department of Mechanical Engineering, Indian Institute of Technology Madras, 600036, India}

\begin{abstract} 
The water entry of solid bodies, in particular spheres, has been the subject of numerous studies due to its relevance in multiple applications, especially focusing on cavity dynamics and the resulting forces and subsequent trajectories. The role of wettability in cavity formation and sustainability has been experimentally and theoretically addressed in the past; however, typically only up to the occurrence of a primary air cavity pinch-off. What is less explored is how wettability can influence higher-order pinch-offs. In the following study, especially secondary pinch-offs are shown to determine whether a sphere trajectory is terminally ascending, i.e., the sphere returns to the surface, or terminally descending: wettability is shown to be a strong influencing factor.

This study begins with a series of systematic experiments, comparing the entry and trajectory of spheres over a range of impact Reynolds and Froude numbers, but also with varying surface wettability. From time resolved trajectories, a force balance is computed, illustrating the role of buoyancy, arising from the volume of any sustained air cavity in the wake of the sphere. Secondary pinch-offs alter this volume; hence, the terminal trajectory. A force-scale comparison and local radial balance at the cavity neck are then presented, suggesting how wettability influences the secondary pinch-off.

\end{abstract}


\maketitle
\section{Introduction}

The entry of a solid object into a liquid pool, commonly referred to as the water-entry problem, is a fundamental phenomenon in fluid mechanics with broad relevance across industrial, athletic, and biological contexts. These problems involve air entrainment, cavity formation, and transient hydrodynamic forces, which collectively govern the subsequent motion and trajectory of the impacting body. Understanding these interactions is crucial for a wide range of applications, including marine engineering and seaplane landings \cite{seddon2006review}, vertical missile water entry \cite{May1952VerticalWater}, projectile-water impact analysis \cite{Truscott2014WaterProjectiles} and underwater robotics \cite{yuh2000design, neira2021review}. A recent review by \citet{jung2025water} highlighted the shared strategies governing water-entry and exit phenomena in both biological and engineering systems, demonstrating how insights into impact forces, body morphology, and dynamics can inspire advanced marine and aerial–aquatic designs.

The entry and penetration of a sphere into a deep liquid pool serves as a simplified yet generic representation of these practical applications. The foundational work in this area was conducted by \citet{worthington1883impact}, who used high-speed photography to pioneer the study of a solid sphere impacting on the liquid. Subsequent research extended these insights, examining the influence of air pressure on splash and cavity formation \cite{Gilbarg1948InfluenceWater}, potential flow based models of cavity geometry \cite{richardson1948impact}, the effects of surface roughness and impact angle on cavity dynamics \cite{may1965study, laverty2005experimental}. Later studies further explored the influence of surface grooves and impact velocity on cavity behavior \cite{Mehri2021WaterVelocity}, as well as the effects of sphere heating and the formation of Leidenfrost vapor layers on sphere dynamics \cite{Mehri2020HydrodynamicEntry,Vakarelski2014LeidenfrostWater,Mansoor2017Stable-streamlinedSpheres,Vakarelski2019Stable-streamlinedWater}. Further experimental and numerical investigations have examined the unsteady motion of hydrophilic and hydrophobic spheres, including pinch-off dynamics, vortex formation, and the resulting hydrodynamic forces \cite{Duclaux2007DynamicsCavities,Aristoff2009WaterSpheres,aristoff2008water,Truscott2012UnsteadyEntry,vakarelski2011drag,Speirs2019WaterAnglesb,li2019experimental}. In addition, the effects of stratified liquids \cite{tan2016experimental}, as well as capillary and viscous forces \cite{Marston2011BubbleSurfaces}, have been extensively investigated. The reduction of drag during underwater motion has also been the subject of recent studies \cite{zhu2022sustaining,zhao2025bioinspired}. These studies have substantially advanced our understanding of cavity formation, sphere encapsulation, and deceleration processes. However, the influence of post-pinch-off air cavities on the subsequent dynamics of the sphere remains relatively underexplored.

Surface wettability has emerged as a key factor governing water-entry dynamics. Hydrophobic and  superhydrophobic surfaces can significantly alter wetting behavior, stabilize air layers, and influence cavity growth, entrained air volumes, and post-impact motion \cite{Truscott2006CavitySurface,Duclaux2007DynamicsCavities,Mansoor2014WaterFormation,Vakarelski2019Mobile-surfaceStronger,Guleria2021ExperimentalSphere}. During the initial stages of impact, hydrophobic spheres typically generate an air cavity that undergoes pinch-off, leaving a residual bubble or air cavity in the wake of the sphere \cite{Aristoff2009WaterSpheres,Truscott2012UnsteadyEntry,Vakarelski2017Self-determinedCavities}. The size and persistence of this wake cavity depend on several factors, including the impact Reynolds number, density ratio, surface wettability, impact angle \cite{asfar1987rigid}, and possible rotation of the sphere \cite{techet2011water}. \citet{truscott2009cavity} and \citet{Truscott2014WaterProjectiles,Truscott2012UnsteadyEntry} further demonstrated that variations in contact angle, sphere density, and impact velocity significantly affect vortex formation and the resulting unsteady hydrodynamic forces during penetration, as revealed through Particle Image Velocimetry (PIV) measurements.

Previous investigations on sub-millimeter superhydrophobic (SH) spheres have identified distinct regimes of bouncing, oscillation, and sinking, governed by Weber and Bond numbers, surface energy, and capillary effects \cite{Maru1971BehaviorInterface,Galeano-Rios2017Non-wettingDroplets,Wang2015BehaviorDropletsurface,Chen2018EntrappingInterface}. The water entry of SH cylinders was examined by~\citet{Liu2007BuoyantWater}, employing potential flow and hydrodynamic theories to describe the sinking cylinder motion. Subsequently, capillary dominated impacts were investigated by~\cite{Vella2006TheObjects,Vella2007SurfaceImpact,vella2015floating}, delineating the conditions under which a cylinder transitions from floating to sinking in fluids. In the axisymmetric case,~\citet{Lee2008ImpactWater} studied the impact of small SH spheres on liquid surfaces and proposed scaling laws to predict transitions among sinking, oscillation, and bouncing regimes based on capillary forces over a wide range of Weber and Bond numbers. A mathematical model was also derived to describe the early and late stages of rebound dynamics of the sphere after impacting the liquid.
 
Despite extensive prior research, the role of post-pinch-off air cavities, especially secondary pinch-offs, and their influence on the subsequent motion of impacting spheres remain inadequately understood, particularly for lighter spheres ($\rho_s < 2000 $ kg/m$^3$). This knowledge gap is particularly critical for superhydrophobic spheres of relatively low density, where entrained air volumes can substantially modify drag and buoyancy forces, often leading to unexpected terminal motions. The present study addresses this gap through a systematic investigation covering a wide range of sphere diameters and impact velocities. Using high-speed imaging and image based analysis, this study quantifies the instantaneous trajectories, velocities, and accelerations of superhydrophobic spheres through spline fitting and smoothing techniques, elucidating how the total displaced volume and air cavity dynamics govern the transition between terminally descending and ascending motion, depending on the primary and secondary pinch-off events. The results indicate that abrupt changes in the entrained air volume during primary, secondary, and sometimes tertiary pinch-off lead to corresponding variations in buoyancy force, ultimately determining whether the sphere undergoes terminally descending or ascending motion. Preliminary observations by the authors have revealed a one-to-one correlation between the result `terminally ascending' and whether a secondary pinch-off of the attached air cavity is avoided. The motivating question can therefore be narrowed to `what factors influence higher-order pinch-off events?'

 The phenomenon of higher order pinch-offs and their influencing factors was first approached experimentally using high-speed visualization. This allowed a narrower parametric window to be established, within which the quantitative influence of various factors could be independently studied. This was followed by systematic parameter variations which capture the primary influences of pinch-off, admittedly allowing for some uncertainty in determining some of the input parameters. This includes the exact volume of the air cavity, as only projections are available from the visualizations. Another uncertain factor is the instantaneous contact angle and position of the three-phase contact line defining the wake cavity. Understandably, this angle cannot be accurately measured or directly related to contact angles measured in a stationary state, as with a sessile drop.

\section{\label{sec:experimental details}Experimental apparatus and image analysis}
 
\subsection{Experimental set-up}

\begin{figure}[ht!]
\centering
\includegraphics[width=0.85\textwidth]{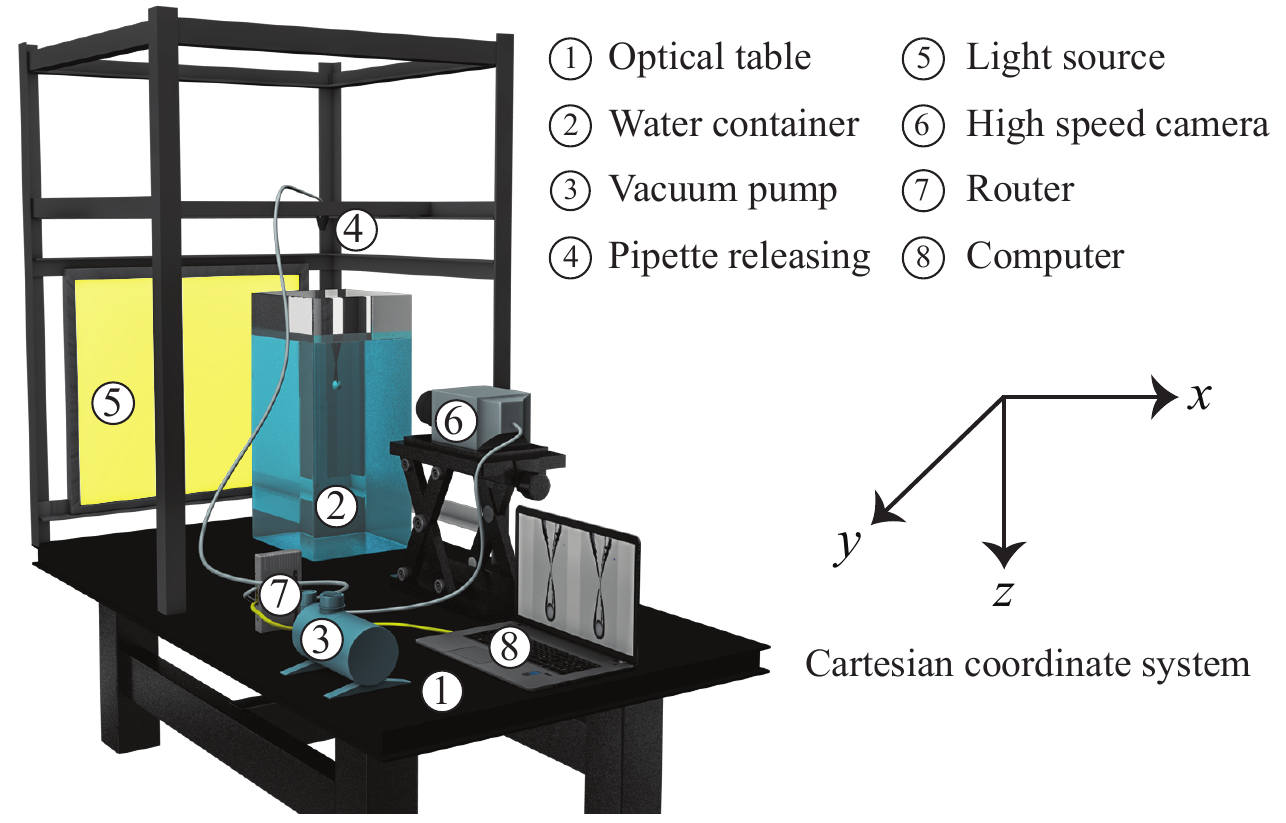}
\caption{Schematic of the experimental setup, showing a water container of dimensions 200 (L) $\times$ 200 (B) $\times$ 400 (H) mm$^3$. Spheres of diameters 4 mm, 6 mm, and 10 mm are released from controlled heights using a vacuum pump. } 
\label{fig:exp_setup}
\end{figure}

The experimental setup, illustrated in Fig.~\ref{fig:exp_setup} and similar to that in Ref.~\cite{Billa_Josyula_Tropea_Mahapatra_2025}, consists of a transparent acrylic container, a high-speed monochrome camera, a backlight, a computer to record the motion, and a vacuum based release system. The container has a cross-section of 200 (L) $\times$ 200 (B) $\mathrm{mm}^2$, which is 20 times the diameter of the largest sphere (10~mm) used in this study, ensuring minimal wall effects on the trajectories of the sphere. The depth of the container is 400~mm. The camera (Phantom VEO E-340-L equipped with a Nikon 24-85~mm microlens) records at 1000~fps with a resolution of 896~$\times$~1600 pixels and an exposure time of 90~$\mu$s. The field of view of the camera is 100~mm~$\times$~150~mm. For experiments focused on elucidating interface dynamics, a second high-speed camera (Phantom VEO E-340-L equipped with a Nikon 28~mm microlens) operates at 2000~fps with a resolution of 1024~$\times$~1024~pixels and an exposure time of 90~$\mu$s, capturing the moment of impact and the subsequent evolution of the cavities. To ensure consistent results, every experiment was conducted at least three times, allowing a 15 minute interval between trials to minimize air-water fluctuations, meniscus distortions, and water settling.

The vacuum release system comprises a pipette tip, a connecting pipe, and a vacuum pump, which generates suction pressure at the pipette tip to securely hold the spheres. The spheres are released by deactivating the suction pressure. All experiments were conducted in a controlled indoor environment at an ambient temperature of 25$^\circ$C. It is ensured that there is no rotation or lateral deviation of the sphere before impacting the free air-water surface.

Spheres were made superhydrophobic (SH) using a commercially available coating (NeverWet spray). The spray is a two-step system designed to create a moisture repelling barrier on the substrate, as further described in \cite{billa2026entry}. 

\subsection{Experimental methodology}
\label{subsec:experimental methodology}
\begin{table}[b]
\caption{\label{tab:parameters}
Properties of hydrophobic and superhydrophobic spheres with associated range of values.
}
\begin{ruledtabular}
\begin{tabular}{c c c c }
\textrm{Parameter}&
\textrm{Symbol}&
\textrm{Definition}&
\textrm{Range of values}\\
\colrule
        Sphere diameter & $D$ & -  & 4, 6, 10~mm  \\
      Density ratio &$\rho^*$ & $\rho_s/\rho_l$  & 1.12, 1.37, 2.16, 3.26, 6.08, 7.92  \\
       Impact velocity & $v_i$ & $ \sqrt{2gh} $  & 0.89 - 3.52~m/s \\
     Impact Froude number & $\mathrm{Fr}_i$ & $ v_i/\sqrt{gD}$  & 2.84 - 17.77  \\
\end{tabular}
\end{ruledtabular}
\end{table}

The impact conditions of hydrophobic and superhydrophobic spheres, together with the relevant parameters and dimensionless quantities, are summarized in Table~\ref{tab:parameters}. The quasistatic advancing contact angle of deionized (DI) water on the sphere surface ($\theta_a$) was measured using the sessile drop method at multiple locations while accounting for surface curvature effects. With this method, the volume of the sessile drop is increased until the contact line just moves, and this angle is denoted the quasistatic advancing contact angle. For the hydrophobic sphere, the contact angle was determined using the ellipse-fit method in ImageJ, yielding an average value of $\theta_a =100 \pm 3^{\circ}$. During measurement, the water drop was placed such that the $z$-axis passed through the apex of the drop.  Following the procedure described by \cite{Extrand2012IndirectSurfaces, Extrand2002WaterSurfaces, Magos2021ContactSurfaces,Viswanadam2012ContactSurfaces}, the apparent angle was obtained using trigonometric relations between the chord half-length ($b$) and the radius of the sphere ($R_0$), accounting for the local curvature of the surface (see Fig.~\ref{fig:TPCL_CAA}(a)). In the case of the superhydrophobic sphere, direct measurement on the curved surface was difficult; therefore, the contact angle was measured on a flat plate coated with the same superhydrophobic spray (NeverWet), giving $155 \pm 2^{\circ}$. As a comparison, \citet{weisensee2016water} performed similar measurements on flat substrates (glass: 2700~kg/m\textsuperscript{3}, and polymethylmethacrylate: 1180~kg/m\textsuperscript{3}) coated with NeverWet spray, and reported advancing and receding angles of $164 \pm 4^{\circ}$ and $159 \pm 3^{\circ}$, respectively. Angles exceeding $150^\circ$ are generally considered as superhydrophobic.

Sphere diameters ($D$) were measured using a Vernier caliper, and masses ($m$) with an electronic balance (Ohaus), yielding the density ($\rho_s$) and the corresponding density ratio ($\rho^* = \rho_s / \rho_l$, with $\rho_l = 1000$ kg/m$^3$). The diameters of the impacting spheres were 4~mm, 6~mm, and 10~mm, released from controlled heights ($h$) above the water surface in the range $49~\text{mm} \leq h \leq 633~\text{mm}$, resulting in impact velocities ($v_i \approx \sqrt{2gh}$)  in the range $0.89~\text{m/s} \leq v_i \leq 3.52~\text{m/s}$. The Froude number, $\mathrm{Fr_i} = v_i/\sqrt{gD}$, ranged between 2.84 and 17.77. Here, $g = 9.81~\mathrm{m/s^2}$ is the acceleration due to gravity. In the discussions below, length scales are made dimensionless using the diameter of the sphere $D$, velocities using the impact velocity $v_i$, displaced volumes with the sphere volume $V_S = \pi D^3/6$, and time using the characteristic scale $D/v_i$. All variables made dimensionless are indicated by a superscript $(^{*})$.

\subsection{Image analysis}
\label{subsec:image analysis}

\begin{figure}[ht!]
\centering
\includegraphics[width=\textwidth]{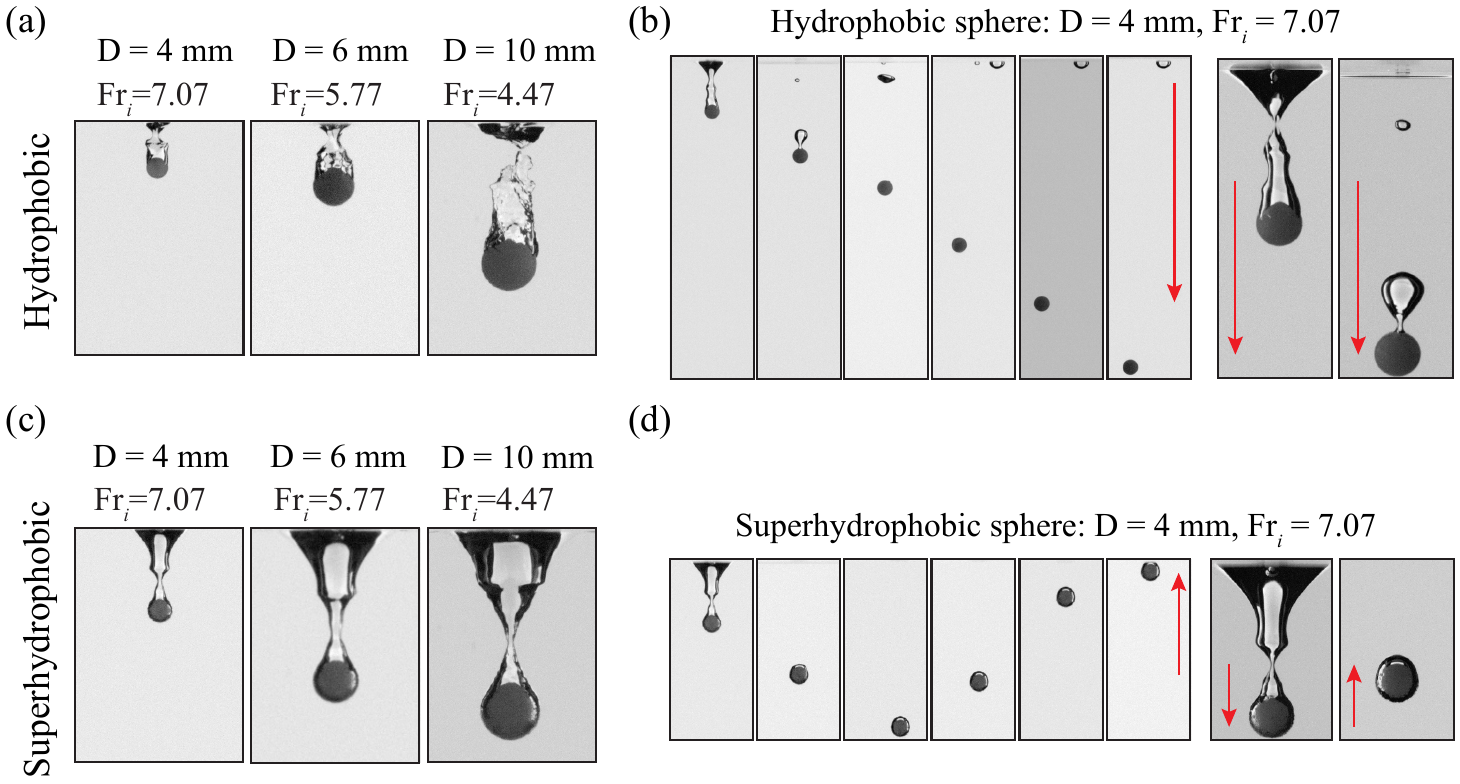}
\caption{Images illustrating cavity formation and the motion of hydrophobic and superhydrophobic (SH) spheres during water entry: (a) Air cavity formation for hydrophobic spheres (4, 6, and 10 mm; $\mathrm{Fr}_i = 7.07$, 5.77, and 4.47, respectively) during penetration. (b) Sequential images showing the terminally descending motion of a 4 mm hydrophobic sphere at $\mathrm{Fr}_i = 7.07$ with complete cavity rupture. (c) Air cavity formation for SH spheres under identical conditions of (a). (d) Sequential images showing the terminally ascending motion of a 4 mm SH sphere at $\mathrm{Fr}_i = 7.07$ with a persistent air-layer. The last two frames in (b) and (d) show enlarged views highlighting regions near and after pinch-off. The red arrow denotes the direction of motion.} 
\label{fig:Nc_SH_raw}
\end{figure}

Experimental data consist of video sequences in which the trajectory of the sphere can be followed over time, as illustrated in Fig.~\ref{fig:Nc_SH_raw} for entry of hydrophobic and superhydrophobic spheres of varying diameter and impact Froude number. The hydrophobic sphere, shown in  Figs.~\ref{fig:Nc_SH_raw}(a) and (b) exhibits a relatively large air cavity following primary pinch-off, with unsteady cavity walls and a three-phase contact line on the sphere close to the equator. However, for the 4~mm sphere, a second pinch-off occurs shortly after, as illustrated in the inset shown in Fig.~\ref{fig:Nc_SH_raw}(b) and the sphere continues to descend in the pool.

In contrast, following primary pinch-off the superhydrophobic sphere exhibits a more stable air cavity with smooth sidewalls and a three-phase contact line much closer to the north pole of the sphere (Fig.~\ref{fig:Nc_SH_raw}(c)) than with hydrophobic spheres. As the inset in Fig.~\ref{fig:Nc_SH_raw}(d) shows, no further pinch-off occurs for the 4~mm sphere, and the sphere eventually reverses direction and ascends back to the surface of the pool. These two examples illustrate the influence of wettability on the sustainability of the air cavity; hence, on the terminal motion behavior of the sphere. The explanation of this role of wettability is central to the present study.

\begin{figure}[ht!]
\centering
\includegraphics[width=0.9\textwidth]{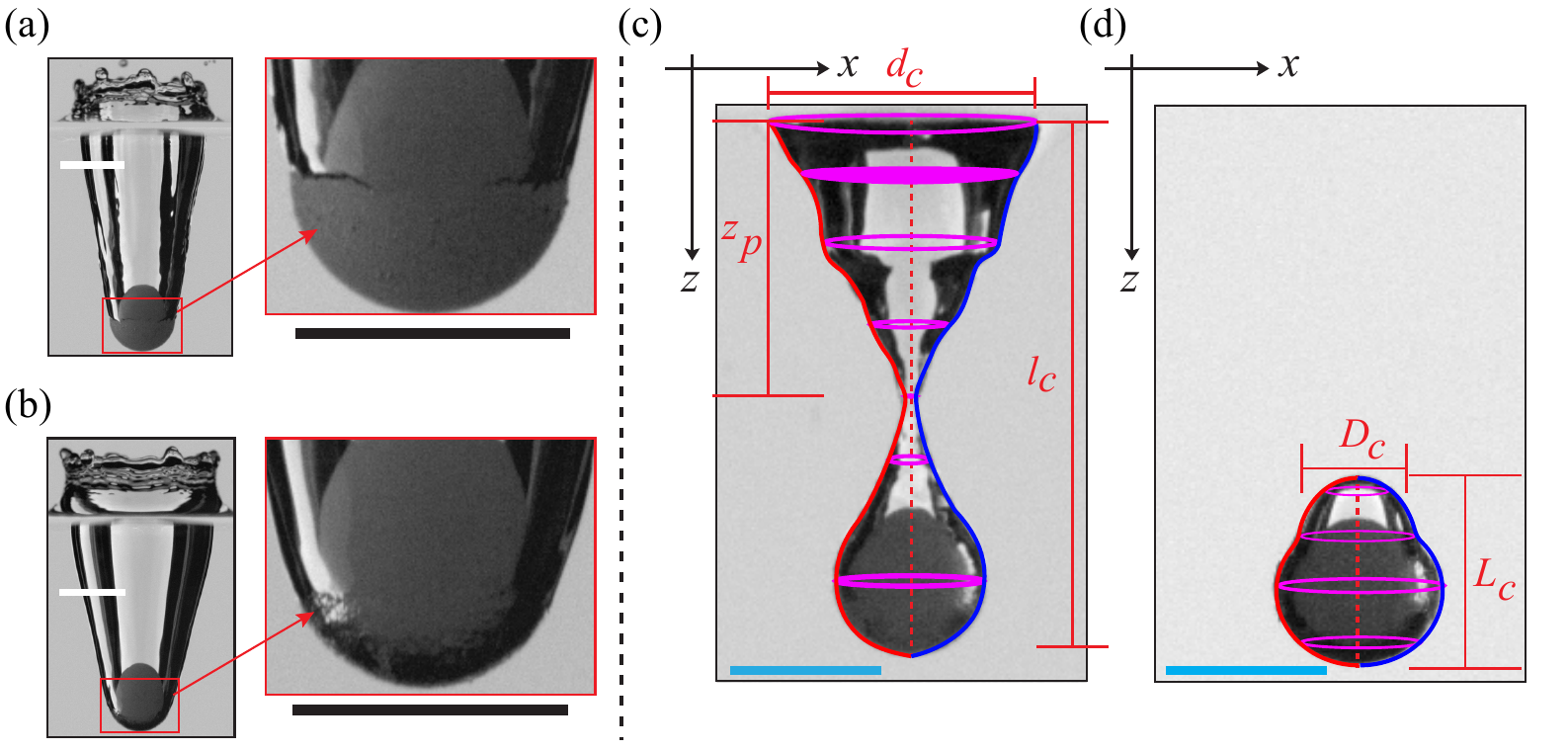}
\caption{Images of hydrophobic and superhydrophobic spheres during water entry. (a) Hydrophobic sphere showing cavity formation and complete wetting up to the equator (inset). (b) Superhydrophobic sphere exhibiting smooth cavity formation with complete air encapsulation of the sphere during penetration (inset). (c,d) Schematic illustration of how the displaced volume was computed following~\cite{billa2026entry}, where the cavity profile is divided into horizontal disks and integrated along the vertical direction using the trapezoidal method. Panel (d) shows a sustained bubble-tail cavity with the associated geometrical parameters. Scale bars in all images and insets represent 10~mm (equal to the sphere diameter).} 
\label{fig:TPCL_CAA}
\end{figure}

Anticipating the need to estimate the displaced volume, further image processing was performed, as illustrated in Fig.~\ref{fig:TPCL_CAA}. Images in Fig.~\ref{fig:TPCL_CAA}(a) and (b) show the air cavity prior to primary pinch-off for a hydrophobic and superhydrophobic sphere, respectively. While the hydrophobic sphere exhibits a three-phase contact line near the equator, the superhydrophobic sphere is completely encapsulated in an air pocket. The computation of the displaced volume is illustrated in Fig.~\ref{fig:TPCL_CAA}(c) and (d) just before and after primary pinch-off. The displaced volume was computed at each time step following the procedure of \cite{billa2026entry}, using edge detection combined with trapezoidal integration of horizontal disks in the vertical direction. The total displaced volume was then made dimensionless using the volume of the sphere, $V_S$, yielding the dimensionless displaced volume ($V_\mathrm{Disp}^* = V_\mathrm{Disp}/V_S$). 

In processing the images to obtain either sphere position or dimensions of the displaced volume from the high-speed videos, the position uncertainty was estimated as follows. Knowing the measured sphere diameters, the magnification could be calibrated, resulting in a pixel size $\Delta_{\mathrm{px}} = 0.059$~mm at 2000~fps and $\Delta_{\mathrm{px}} = 0.145$~mm at 1000~fps. Interfaces were found by edge detection, typically blurred over two pixels. Thus, the uncertainty in edges is considered to be $\pm 1$~pixel. A radius measured as half the distance between two such edges, which are independent, is then uncertain by $\delta r = \pm\Delta_{\mathrm{px}}/\sqrt{2}$. For the neck radius $R_n$, i.e.,\ the smallest radius of the attached cavity (Fig.~\ref{fig:fbd_Forces}), normalized by the sphere radius $R_s = D/2$, this gives $\delta R_n/R_s \approx 0.05$, $0.04$ and $0.03$ for $D = 4$, $6$ and $10$~mm, respectively. The displaced volume is obtained by summing horizontal disks of thickness $\Delta z$ and local radius $r(z)$, $V_{\mathrm{Disp}} = \sum \pi r^2 \Delta z$~\cite{billa2026entry}, so that an error $\delta r$ in the radius changes the volume of the attached air, $V_{\mathrm{air}} = V_{\mathrm{Disp}} - V_S$, by a fraction $2\,\delta r/\bar r$, where $\bar r$ is the mean radius of the cavity. For a 4~mm sphere, $\bar r \approx 1.1$~mm and $\delta r \approx 0.10$~mm give about 18\% of $V_{\mathrm{air}}$; since the sphere volume is known exactly, this translates into about 11\% uncertainty in $V_\mathrm{Disp}$ for the 4~mm spheres and about 4\% for the 10~mm spheres. The axisymmetry assumption was checked by revolving the left and right halves of each edge profile separately about the cavity axis, obtained by a straight-line fit through the mid-points of the profile to allow for lateral offset and tilt. Before secondary pinch-off, the two volumes differ by 1.4\% on average for D = 4 mm and 4.1\% for D = 6 mm (maximum 3.4\% and 8.1\%), smaller than the edge detection uncertainty. This test is restricted to the image plane; it varies from rotational symmetry in the viewing direction, such as an elliptical cross section, and cannot be detected with a single camera, and remains unquantified. The volume itself is obtained from horizontal disks with a radius equal to half the local width, so the lateral offset of the cavity from the sphere centre (0.3--0.9~mm) does not affect it. There is no correction for the inclination angle $\beta$ of the cavity axis to the vertical (3--5$^\circ$). For a circular cavity inclined in the image plane, it overestimates the volume by a factor $1/\cos\beta$, i.e., \ by less than 0.4\%. 

After the primary pinch-off, no air detaches from the cavity until the secondary pinch-off, and the displaced volume remains constant. The gas volume could still change with pressure, but the hydrostatic pressure changes by only about 0.1\% of atmospheric pressure per 10~mm of depth, and the capillary pressure $2\sigma/R_n$ remains below 1\% of it, so that compression of the gas is negligible. When a single value of the volume is needed, a mean value is used as an approximation, averaged over all frames from the first frame in which a neck is visible to the last frame before the secondary pinch-off for descending spheres.

Events occurring between two consecutive frames are assigned to the midpoint of the frame interval, with an uncertainty of $\pm \Delta t/2$. This convention applies to the impact time ($t = 0$), the onset of necking ($t_\mathrm{n}$), and the secondary pinch-off time ($t_\mathrm{2p}$). At 1000 fps, an uncertainty of $\pm$0.5~ms corresponds to $\pm 0.04 - 0.44$ in $t^*$, which is negligible compared with the duration of the trajectories ($t^*\sim 10^2 - 10^3$). Near pinch-off, however, the relevant time scale is the remaining time $t_{2p} - t$ rather than the duration of the trajectory. With the neck evaluated every 0.5--2~ms, $t_{2p}$ is uncertain by $\pm0.25$--$1$~ms, which is 25--100\% of a remaining time of 1~ms. The points closest to pinch-off in Fig.~\ref{fig:neck_collapse}(b), shaded in the figure, therefore carry a large uncertainty in $t_{2p} - t$.

\subsection{Forces and equations of motion}
\label{subsec:ge}
The motion of the sphere along its trajectory is governed by body forces, apparent forces, hydrodynamic forces, and inertial forces. The body forces consist of the gravitational force ($F_\text{G}$ - weight) and the buoyancy force ($F_\text{B}$). Apparent forces include the added mass force ($F_{\text{AM}}$) and the Basset history force ($F_\text{H}$), which account for inertial and unsteady viscous effects respectively. The hydrodynamic drag force ($F_\text{D}$) comprises both viscous and pressure contributions, as well as capillary forces arising from surface tension ($F_\text{S}$). The inertial force ($F_\text{I}$) accounts for the acceleration/deceleration of the sphere.

For the experiments performed in this study, the sphere follows a nearly vertical path during its motion; thus, lift forces are neglected. Examples of water entry of hydrophilic spheres exhibiting significant lift forces, their computation from trajectory data, and the corresponding non-vertical trajectories can be found in \cite{Billa_Josyula_Tropea_Mahapatra_2025}. Previous experiments have documented that the sphere experiences relatively low acceleration following the primary pinch-off (as shown below in Fig.~\ref{fig:velocity_acceleration}(d)), allowing both of the apparent forces to be eliminated from the force balance. Moreover, once there is no longer an uninterrupted interface to the free surface, i.e., after primary pinch-off, there are no net capillary forces acting on the sphere plus cavity in the direction of motion. The scalar momentum equation along the $z$-axis and post-pinch-off then reads
\begin{equation}
      \underbrace{\frac{1}{6} \rho_s \pi D^3 \frac{\mathrm{d}v_\mathrm{s}(t)}{\mathrm{d}t}}_\text{Inertial force}=  
      \underbrace{\frac{1}{6} \rho_sg\pi D^3}_\text{Gravity}-
      \underbrace{\rho_lgV_\mathrm{Disp}(t) }_\text{Buoyancy}        - \underbrace{\frac{1}{8}\rho_l C_D  \pi D^2 v_{s}(t) |v_{s}(t)|}_\text{Drag}
       \label{eq:Force_evaluation} 
\end{equation}
\noindent where $v_s$ is the velocity of the sphere and $C_D$ is the drag coefficient. In the non-accelerating state, the drag coefficient is taken as 0.5, which was also directly measured in previous experiments \citep{Billa_Josyula_Tropea_Mahapatra_2025} and agrees well with literature values for the Newtonian range of Reynolds numbers encountered in this experiment. Although the coordinate system is chosen such that positive $z$ is in the downward direction, the sign of the velocity ($v_s$) and of each force is chosen positive in the direction of motion; thus, if a sphere transitions from descending to ascending, the signs of the forces would change. Any sphere that decelerates, changes direction, and returns to the surface must exhibit a buoyancy force larger than the gravitational force, expressed as
\begin{equation}
    \label{eq:condition for ascending}
    \frac{F_\mathrm{B}}{F_\mathrm{G}}=\frac{\rho_\mathrm{l} g V_\mathrm{Disp}}{m_\mathrm{s}g}=
    \frac{V^*_\mathrm{Disp}}{\rho^*}>1 \quad \quad \mathrm{or} \quad  \quad V^*_\mathrm{Disp}>\rho^*
\end{equation}
thus, whether a sphere 'terminally ascends' or 'terminally descends' depends on the size of the air cavity, if and once the sphere decelerates to zero velocity.

The dimensionless terminal velocity for a fixed air cavity volume is obtained by equating the net body forces with the hydrodynamic force (from Eq.~(\ref{eq:Force_evaluation})), leading to
\begin{equation}
v_t^* =
\frac{\operatorname{sgn}\left(\rho^* - V_{\mathrm{Disp}}^*(t)\right)}{\mathrm{Fr}_i}
\sqrt{\frac{4\left|\rho^* - V_{\mathrm{Disp}}^*(t)\right|}{3C_D}}.
\label{eq:terminal velocity}
\end{equation}
which can be computed knowing $V_{\mathrm{Disp}}^*(t)$ from the images and assuming a value for the drag coefficient.

\begin{figure}
    \centering
    \includegraphics[width=0.9\textwidth]{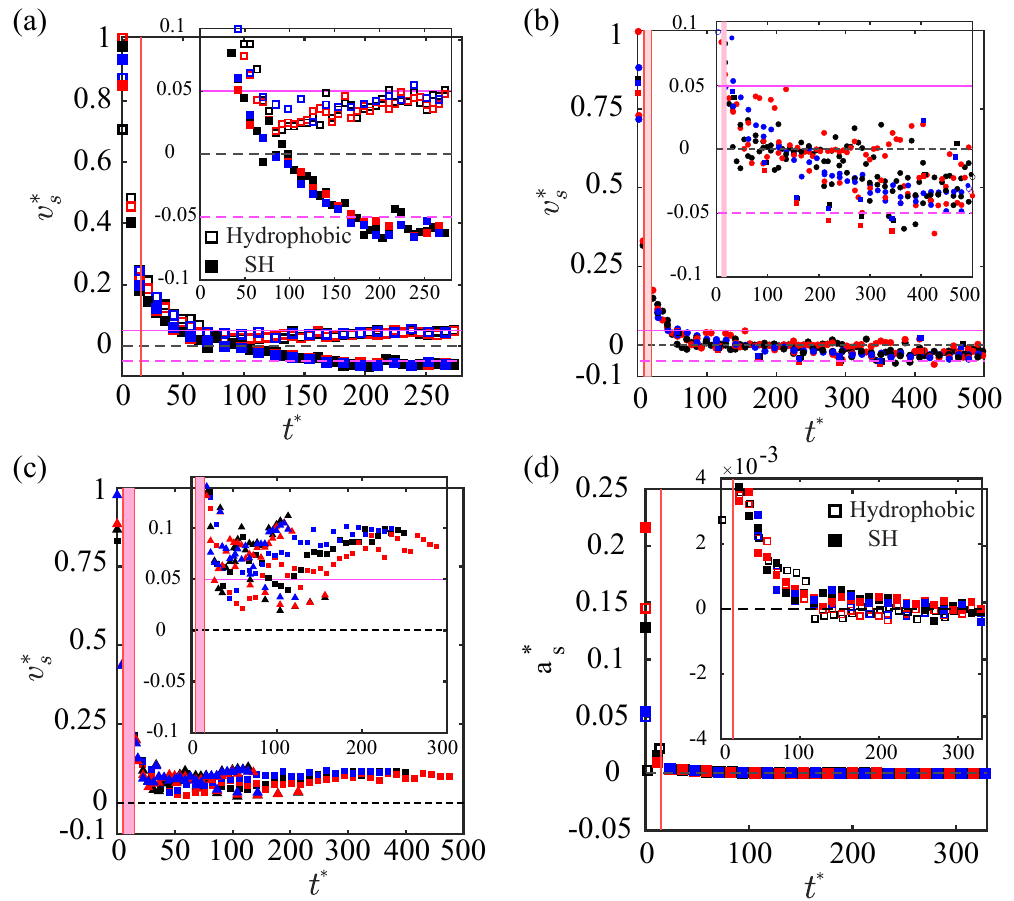}
    \caption{Instantaneous dimensionless velocity and acceleration as functions of dimensionless time for spheres with $\rho^* = 1.12$. (a) Comparison of hydrophobic and superhydrophobic 6 mm spheres at $\mathrm{Fr}_i = 11.55$. (b) Representative terminally ascending cases. (c) Representative terminally descending cases. (d) Corresponding acceleration evolution for hydrophobic and superhydrophobic spheres. Colors denote repeated experiments; markers indicate sphere diameter (circle: 4 mm, square: 6 mm, triangle: 10 mm). The red vertical lines in panels (a) and (d) indicate the dimensionless pinch-off time, while the red shaded regions in panels (b) and (c) represent the dimensionless pinch-off time interval.}
    \label{fig:velocity_acceleration}
\end{figure}

From the high-speed images, the temporal displacement of the sphere's south pole was tracked throughout its entire trajectory within the field of view. The signed vertical velocity  $v_s(t) $ was computed by differentiating this displacement data after first fitting with a quintic spline function.  Dimensionless acceleration ($a_s^*(t)$) was computed from the time derivative of the velocity. To reduce noise in the acceleration data, the velocity was first smoothed using a piecewise spline fit applied over fixed time intervals. The resulting values of $a_s^*(t)$ were further smoothed using a smoothing spline, with a smoothing parameter close to unity \cite{epps2010evaluating}. In this manner, all terms in Eqs.~(\ref{eq:Force_evaluation}) and (\ref{eq:condition for ascending}) could be computed for all times.

To illustrate this procedure, Figs.~\ref{fig:velocity_acceleration}(a) and (d) show the velocity ($v_s^*(t)$) and acceleration data ($a_s^*(t)$) against dimensionless time $t^*$, respectively, for 6~mm spheres released at $\mathrm{Fr}_i = 11.55$, comparing hydrophobic and superhydrophobic spheres. In the hydrophobic case, the cavity undergoes a complete pinch-off, resulting in a reduced buoyancy force. As a result, the sphere continues descending, eventually attaining a constant terminal velocity. In these plots, the terminal velocity (Eq.~(\ref{eq:terminal velocity})) is represented by the magenta dashed line for the ascending case and by the continuous magenta line for the descending case. In contrast, the superhydrophobic sphere retains a cavity, which maintains a higher buoyancy force.  The sphere subsequently accelerates upward until it attains a constant terminal velocity in the ascending direction. In both the terminally ascending and descending scenarios, the acceleration approaches zero as the sphere reaches its terminal velocity, as highlighted in Fig.~\ref{fig:velocity_acceleration}(d). These results of acceleration retroactively justify neglecting the apparent force terms (added mass, Basset history) in Eq.~(\ref{eq:Force_evaluation}) during the  periods of constant velocity.

\section{\label{sec:observations}Experimental Observations}

Preliminary experiments were performed with a selection of different sphere materials of varying density ($1.12 <\rho^*<7.92$) and contact angle ($64^\circ<\theta_a <155^\circ$). These preliminary experiments confirmed two observations:
(a) if a secondary pinch-off occurs, a sphere will not retract to the surface, but continue descending; (b) the parametric range in which terminally ascending behavior is observed is then vastly narrowed, as summarized with the following points:
\begin{itemize}
    \item no sphere of relative density $\rho^* \ge 3.26$ was terminally ascending
    \item no sphere with a measured apparent static contact angle of $\theta_{a}< 97^\circ$ was terminally ascending
    \item superhydrophobic (SH) spheres ($\theta_{a}=155^\circ$) of diameter 4~mm and 6~mm could show terminally ascending behavior.
\end{itemize}
In contrast, we observe that if a secondary pinch-off does not occur, then the buoyancy of the cavity plus sphere after the primary pinch-off is sufficient to counterbalance the force of gravity, i.e., $V^*_\mathrm{Disp}>\rho^*$, and the sphere ascends. One final preliminary observation is that for certain parameter choices, repeated experiments can exhibit both terminally ascending and descending behavior, indicating that there is some residual randomness in the exact behavior of the secondary pinch-off. This occurred in the hydrophobic spheres: two of four experiments ascended for $D = 4$~mm at $v_i = 2.8$~m/s, and two of four for $D = 6$~mm at $v_i = 1.4$~m/s. In all other conditions, every repetition reached the same terminal state.

\begin{table}[ht!]
\caption{Subset of parameters leading to (possible) terminally ascending behavior of spheres.}
\label{tab:parametric_subset}
\begin{ruledtabular}
\begin{tabular}{l c c c}
Parameter & Symbol & Units & Range \\
\hline
Impact velocity & $v_i$ & m/s & 1--3.5 \\
Sphere diameter & $D$ & mm & 4, 6, 10 \\
Density ratio & $\rho^*$ & -- & 1.12, 1.37, 2.16 \\
Apparent static contact angle & $\theta_{a}$ & deg ($^\circ$) & 100, 155 \\
\end{tabular}
\end{ruledtabular}
\end{table}

\begin{figure}[ht!]
\centering
\includegraphics[width=\textwidth]{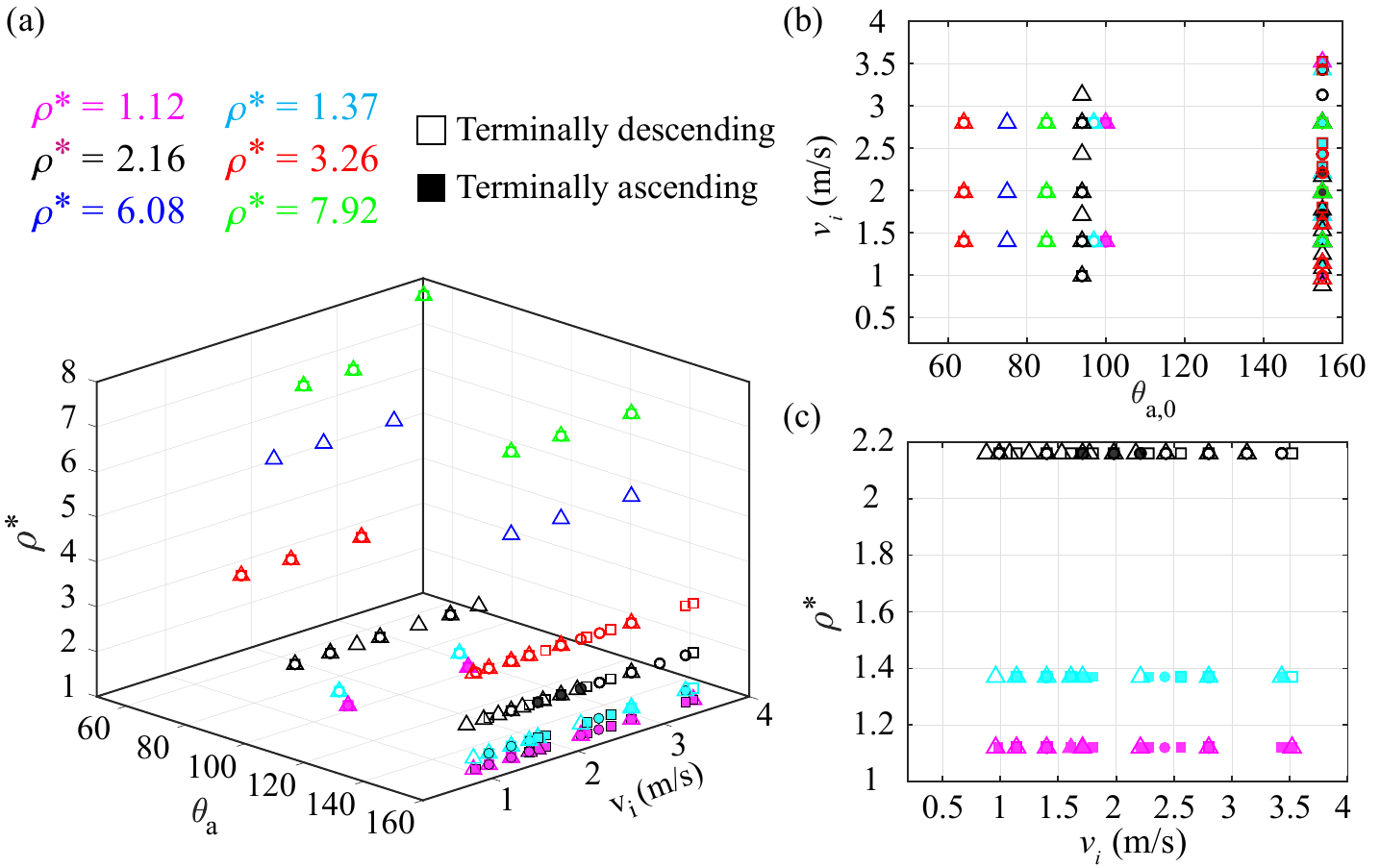}
\caption{(a) Three-dimensional representation of the parameter space showing the variation with density ratio ($\rho^*$), contact angle ($\theta_a$), and impact velocity ($v_i$) over the entire range of conditions. (b) Two-dimensional projection illustrating the variation of $v_i$ with $\theta_a$, where terminal ascending motion is observed only for $\theta_a > 97^\circ$, while all cases below this threshold exhibit terminal descending behavior. (c) Two-dimensional projection showing the variation of $\rho^*$ with $v_i$. Both terminal ascending and descending motions are observed; however, for $\rho^* > 2.16$, all spheres, irrespective of $\theta_a$, $D$, or $v_i$, consistently exhibit terminally descending motion.}
\label{fig:parameter_space} 
\end{figure}

These preliminary observations narrow the parametric range of interest, which is graphically summarized in the three-dimensional plot of Fig.~\ref{fig:parameter_space}(a). In this plot, the terminal behavior of the sphere is shown as a function of surface wettability ($\theta_{a}$), impact velocity ($v_i$), and density ratio ($\rho^*$). In this figure, terminally ascending conditions are marked with filled symbols, and terminally descending with open symbols. Figure~\ref{fig:parameter_space}(b), plotted in the $(\theta_a,\, v_i)$ plane, reveals that terminally ascending cases occur predominantly for highly hydrophobic surfaces with $\theta_a > 97^\circ$. However, the final motion of the sphere is not governed by wettability alone; impact velocity also plays a role through its influence on the secondary pinch-off dynamics. At higher impact velocities,  cavity evolution and subsequent secondary pinch-off reduce the volume of entrained air attached to the sphere, thereby reducing the effective buoyancy force. Consequently, certain cases transition to terminally descending behavior even at $\theta_a \approx 155^\circ$, as indicated by the open symbol markers. In the following section, the primary forces in a local radial balance at the pinch-off are identified and estimated in terms of dimensionless density, impact velocity, and wettability.

The combined influence of impact velocity and density ratio on the terminal motion of the sphere is illustrated in the $(v_i,\, \rho^{*})$ plane (Fig.~\ref{fig:parameter_space}(c)). Spheres with larger density ratios ($\rho^{*} \ge 3.26$) possess greater weight and therefore exhibit a stronger tendency toward terminally descending behavior. In particular, for $\rho^{*} = 2.16$, the $D = 4$~mm spheres exhibit terminally ascending motion under limited conditions, whereas larger spheres predominantly remain descending \cite{billa2026entry}. In contrast, spheres with lower density ratios $(\rho^{*}=1.12$ and $1.37)$ frequently exhibit terminally ascending motion, although the final outcome still depends on the sphere diameter, impact velocity, and the occurrence of secondary pinch-off events.

\begin{figure}[ht!]
\centering
\begin{subfigure}{0.34\textwidth}
    \caption{}
    \centering
    \includegraphics[width=\linewidth]{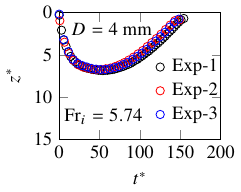}
\end{subfigure}
\hspace{0.001\textwidth}
\begin{subfigure}{0.31\textwidth}
    \caption{}
    \centering
    \includegraphics[width=0.9\linewidth]{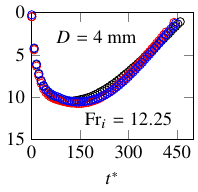}
\end{subfigure}
\hfill
\begin{subfigure}{0.30\textwidth}
    \caption{}
    \centering
    \includegraphics[width=\linewidth]{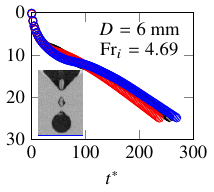}
\end{subfigure}
\begin{subfigure}{0.34\textwidth}
    \caption{}
    \centering
    \includegraphics[width=\linewidth]{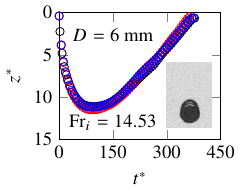}
\end{subfigure}
\hfill
\begin{subfigure}{0.33\textwidth}
    \caption{}
    \centering
    \hspace*{0.5mm}\includegraphics[width=0.92\linewidth]{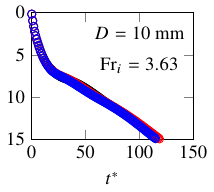}
\end{subfigure}
\hspace{0.001\textwidth}
\begin{subfigure}{0.30\textwidth}
    \caption{}
    \centering
    \hspace*{-4mm}
    \includegraphics[width=0.9\linewidth, trim=1.5 0 0 0,clip]{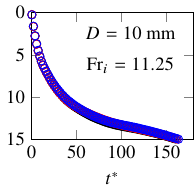}
\end{subfigure}
\caption{ Dimensionless depth ($z^*$) trajectories as a function of dimensionless time ($t^*$) for the lighter superhydrophobic sphere ($\rho^* = 1.12$). (a) $D =$ 4~mm, Fr$_i$~=~5.74; (b) $D =$ 4~mm, Fr$_i$~=~12.25; (c) $D =$ 6~mm, Fr$_i$~=~4.69; (d) $D =$ 6~mm, Fr$_i$~=~14.53; (e) $D =$ 10~mm, Fr$_i$~=~3.63; (f) $D =$ 10~mm, Fr$_i$~=~11.25; }
\label{fig:sphere 1.12}
\end{figure}

Further trajectories are presented in Fig.~\ref{fig:sphere 1.12}, all for superhydrophobic spheres with the lowest dimensionless density ($\rho^*=1.12$) in which both terminal ascending and descending trajectories are found. All of these 4~mm spheres end up ascending, and all 10~mm spheres end up descending, whereas the 6~mm spheres exhibit both behaviors in repeated experiments. When a 6~mm sphere ends up descending, a secondary pinch-off has occurred, lowering the buoyancy force.

\section{Physical interpretation of secondary pinch-off}

Following primary pinch-off, the air retained behind the sphere contributes to the total displaced volume: hence to its buoyancy. Fig.~\ref{fig:Nc_SH_raw} illustrates two outcomes: the attached cavity can undergo a secondary pinch-off, after which the sphere continues to descend, or sufficient air can remain attached for the sphere to reverse direction. Within the reduced force balance of Eq.~(\ref{eq:Force_evaluation}), the buoyancy requirement for upward acceleration at zero sphere velocity is $V_{\mathrm{Disp}}^*>\rho^*$. This condition concerns the motion of the sphere and the attached cavity. It does not determine whether the cavity neck survives.

\begin{figure}[ht!]
\centering
\includegraphics[width=\textwidth]{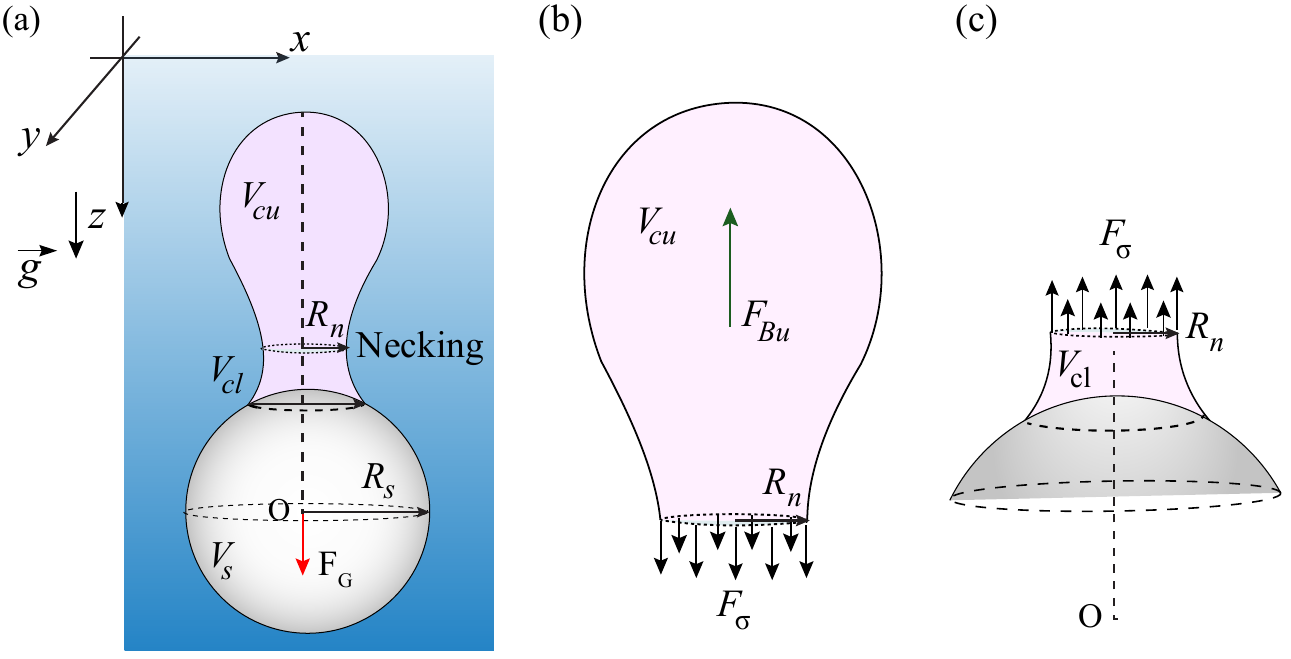}
\caption{(a) Illustration of the geometric parameters of the sphere and air cavity following primary pinch-off, but immediately prior to secondary pinch-off. (b) and (c) Schematic of geometry and selected force contributions in which the air cavity is cut at the neck. Representation of forces acting on the upper (b) and lower (c) sections.}
\label{fig:fbd_Forces}
\end{figure}

The neck geometry and its evolution must therefore be considered separately from the buoyancy requirement. Figure~\ref{fig:fbd_Forces} identifies the upper and lower cavity volumes, $V_{cu}$ and $V_{cl}$, and the minimum neck radius, $R_n(t)$. The following analysis first defines a comparison of body-force and capillary-force scales, and then states the local pressure balance governing radial neck motion. These descriptions have different roles: the force comparison provides a diagnostic for the measurements, whereas the radial balance identifies the additional information needed for a predictive model.

\subsection{Comparison of body-force and capillary-force scales}

The buoyancy of the upper cavity is $F_{Bu}=\rho_l gV_{cu}$, neglecting the weight of the gas. The downward effective weight of the sphere alone is $F_G-F_{Bs}=(\rho_s-\rho_l)gV_S$, where $V_S=4\pi R_s^3/3$. If the lower cavity volume is small compared with the upper cavity volume, $V_{cu}\simeq V_S(V_{\mathrm{Disp}}^*-1)$. A measure of the opposing body-force loads is then
\begin{equation}
 F_{\mathrm{sep}}\equiv F_{Bu}+(F_G-F_{Bs})
 \simeq \rho_l gV_S\Phi,
 \qquad \Phi\equiv V_{\mathrm{Disp}}^*+\rho^*-2.
 \label{eq:iv_load}
\end{equation}
Here $F_{\mathrm{sep}}$ is a load scale, not a derived tensile force transmitted through the neck. A balance for either part of the cavity must also include pressure across the conceptual cut and the relevant hydrodynamic forces and inertia. The sum in Eq.~\eqref{eq:iv_load} does not replace those balances.

For a smooth axisymmetric interface at a local minimum of its radius, the magnitude of the axial surface-tension contribution around the circular neck is $F_\sigma=2\pi R_n\sigma$. Comparing this contribution with Eq.~\eqref{eq:iv_load} gives
\begin{equation}
 \frac{F_{\mathrm{sep}}}{F_\sigma}
 =\frac{2}{3}\frac{R_s}{R_n}\Phi\mathrm{Bo}
 \equiv \frac{R_{\mathrm{ref}}}{R_n},
 \qquad
 \frac{R_{\mathrm{ref}}}{R_s}=\frac{2}{3}\Phi\mathrm{Bo},
 \qquad
\mathrm{Bo}=\frac{\rho_l gR_s^2}{\sigma}.
 \label{eq:iv_reference}
\end{equation}
The reference radius $R_{\mathrm{ref}}$ marks equality of the two selected force scales. Neither $R_n<R_{\mathrm{ref}}$ nor $R_n>R_{\mathrm{ref}}$ is, by itself, a necessary or sufficient condition for pinch-off or cavity retention.

At fixed $\Phi$ and liquid properties, Eq.~\eqref{eq:iv_reference} gives $R_{\mathrm{ref}}/R_s\propto D^2$. However, $\Phi$ depends on the retained volume and need not be the same for different diameters or impact conditions. If the measured volume changes during neck evolution, $\Phi$ and $R_{\mathrm{ref}}$ also change. Comparisons using a mean volume, therefore, represent an average reference level over the specified measurement interval. The uncertainty of the displaced volume enters $R_{\mathrm{ref}}$ through $\Phi = V^*_{\mathrm{Disp}} + \rho^* - 2$. Since $\Phi$ is a small difference of quantities of order unity, $\delta R_{\mathrm{ref}}/R_{\mathrm{ref}} = \delta V^*_{\mathrm{Disp}}/\Phi$ is considerably larger than the relative volume uncertainty: for the experiments in Figs.~\ref{fig:neck_Descending} and~\ref{fig:neck_collapse}, it amounts to 16--21\% for $D = 4$~mm and 12--19\% for $D = 6$~mm. The reference lines are therefore not sharp boundaries. Because $R_n$ and $V^*_{\mathrm{Disp}}$ are obtained from the same profiles, their errors are correlated. Moving all detected edges outward or inward by one pixel and recomputing both quantities changes ($R_{n,\min} - R_{\mathrm{ref}}$)/$R_s$ by at most 0.08 (median 0.04).

\subsection{Local radial motion and interfacial pressure}

A local cylindrical approximation illustrates the pressure information required to describe neck contraction. Assume axisymmetric radial motion in an incompressible, inviscid liquid and neglect axial flow within the local section. The radial velocity is then $u(r,t)=R_n\dot R_n/r$. Let $p_{\ell,n}$ denote the liquid pressure immediately outside the interface and $p_\infty$ the liquid pressure at an outer radial distance $R_\infty$ at the same depth. Integration of the radial Euler equation gives
\begin{equation}
 \rho_l\left[
 \mathcal L\left(R_n\ddot R_n+\dot R_n^2\right)
 -\frac{1}{2}\left(1-\frac{R_n^2}{R_\infty^2}\right)\dot R_n^2
 \right]=p_{\ell,n}-p_\infty,
 \qquad
 \mathcal L=\ln\!\left(\frac{R_\infty}{R_n}\right).
 \label{eq:iv_radial}
\end{equation}
This expression retains the finite outer-radius contribution for the assumed radial velocity field. The outer scale must represent matching to the surrounding flow; it is not an additional measured neck property. For $R_n/R_\infty\ll1$, the term proportional to $R_n^2/R_\infty^2$ can be neglected. Related descriptions of inertial cavity collapse are discussed in \cite{eggers2007theory,gekle2009approach}. Applying this local approximation to the measured minimum radius additionally requires axial transport and motion of the minimum to be negligible over the interval considered.

The liquid pressure $p_{\ell,n}$ must be distinguished from the gas pressure $p_g$. With constant surface tension and negligible viscous normal stresses, the interfacial stress condition is
\begin{equation}
 p_g-p_{\ell,n}=\sigma\kappa,
 \qquad
 \kappa=\frac{1}{a\sqrt{1+a_z^2}}
 -\frac{a_{zz}}{(1+a_z^2)^{3/2}},
 \label{eq:iv_jump}
\end{equation}
where $a(z,t)$ is the local cavity radius and the curvature convention makes $\kappa$ positive for a spherical gas cavity. At a smooth neck minimum, $a=R_n$ and $a_z=0$, so $\kappa=1/R_n-a_{zz}$. The axial curvature therefore reduces the curvature relative to the cylindrical value and cannot be discarded solely because the radius is small. The approximation $\kappa\simeq1/R_n$ requires $|R_na_{zz}|\ll1$.

Combining Eqs.~\eqref{eq:iv_radial} and \eqref{eq:iv_jump}, and taking $R_n/R_\infty\ll1$, yields
\begin{equation}
 \rho_l\mathcal L\left[
 R_n\ddot R_n+
 \left(1-\frac{1}{2\mathcal L}\right)\dot R_n^2
 \right]
 =p_g-p_\infty-\sigma\kappa.
 \label{eq:iv_motion}
\end{equation}
If axial curvature is negligible, the capillary contribution on the right-hand side is $-\sigma/R_n$. The pressure difference $p_g-p_\infty$ remains to be determined. Negligible gas compression does not specify this difference, and dividing the axial load scale $F_{\mathrm{sep}}$ by the neck area does not supply a derived radial-pressure closure. The gas pressure and the surrounding liquid pressure depend on the evolution of the connected cavity and the external flow. Equation~\eqref{eq:iv_motion} is consequently a local balance, not a closed prediction of the measured neck evolution in time.

For a neck instantaneously at rest radially, Eq.~\eqref{eq:iv_motion} gives
\begin{equation}
 \dot R_n=0:\qquad
 \ddot R_n<0
 \quad\Longleftrightarrow\quad
 p_g-p_\infty<\sigma\kappa.
 \label{eq:iv_rest}
\end{equation}
This is a condition for inward acceleration under the local approximation, not a stability criterion. In particular, the formation of a visible neck does not imply that $\dot R_n=0$. For an evolving neck, the corresponding acceleration follows from
\begin{equation}
 R_n\ddot R_n=
 \frac{p_g-p_\infty-\sigma\kappa}{\rho_l\mathcal L}
 -\left(1-\frac{1}{2\mathcal L}\right)\dot R_n^2.
 \label{eq:iv_acceleration}
\end{equation}
Thus, the instantaneous radius alone does not specify the subsequent motion. The radial velocity, pressure difference, curvature, and outer flow also enter. These quantities are not all resolved by the present measurements, and no collapse time or universal critical radius is inferred from Eq.~\eqref{eq:iv_motion}.

\subsection{Interpretation of the measured neck evolution}

\begin{figure}[h]
\centering
\includegraphics[width=\textwidth]{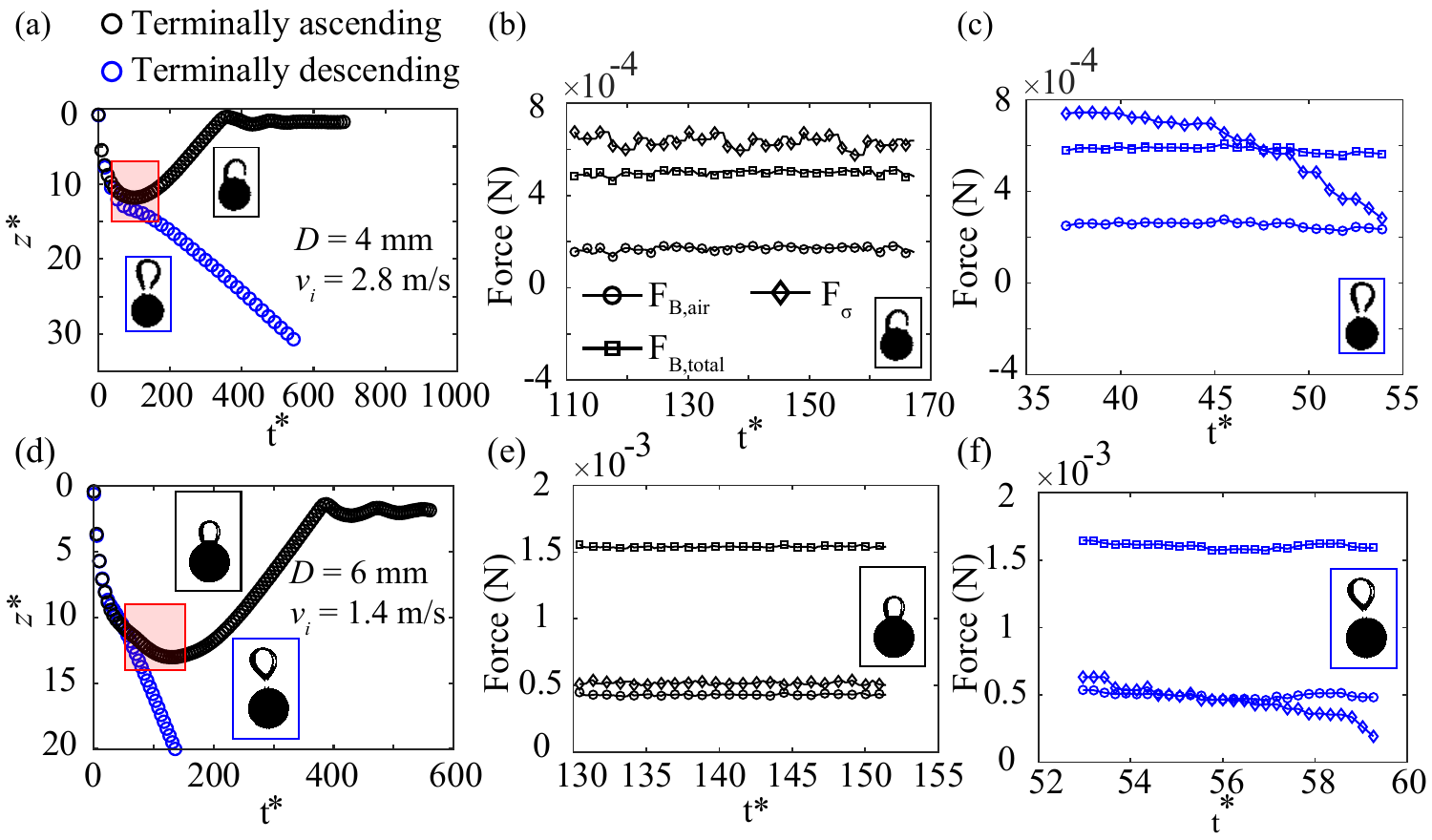}
\caption{Terminal motion of spheres under identical impact conditions ($\rho^*=1.12$ and $\theta_{a}=100^\circ$). Panels (a--c) correspond to $D=4$~mm and $v_i=2.8$~m/s, while panels (d--f) correspond to $D=6$~mm and $v_i=1.4$~m/s. Black and blue symbols denote terminally ascending and terminally descending motions, respectively. The red shaded regions in (a) and (d) indicate the time intervals examined in (b,c) and (e,f), respectively. Panels (b) and (e) show the force evolution for the terminally ascending cases, whereas panels (c) and (f) correspond to the terminally descending cases. The inset images show the final cavity configuration. For terminally ascending motion (black box), the air cavity remains intact without pinch-off, whereas terminally descending motion (blue box) results in cavity pinch-off. } 
\label{fig:forces_bubble}
\end{figure}

Figure~\ref{fig:forces_bubble} compares ascending and descending trajectories at the same nominal hydrophobic surface condition, sphere diameter, and impact velocity. The descending cases exhibit a decreasing neck radius, whereas the ascending cases retain a neck whose radius varies comparatively little over the plotted interval. The comparison demonstrates different cavity histories at the same nominal wettability. The influence of changing wettability is illustrated separately by the hydrophobic and superhydrophobic cases in Fig.~\ref{fig:Nc_SH_raw}.

\begin{figure}[ht!]
\centering
\includegraphics[width=\textwidth]{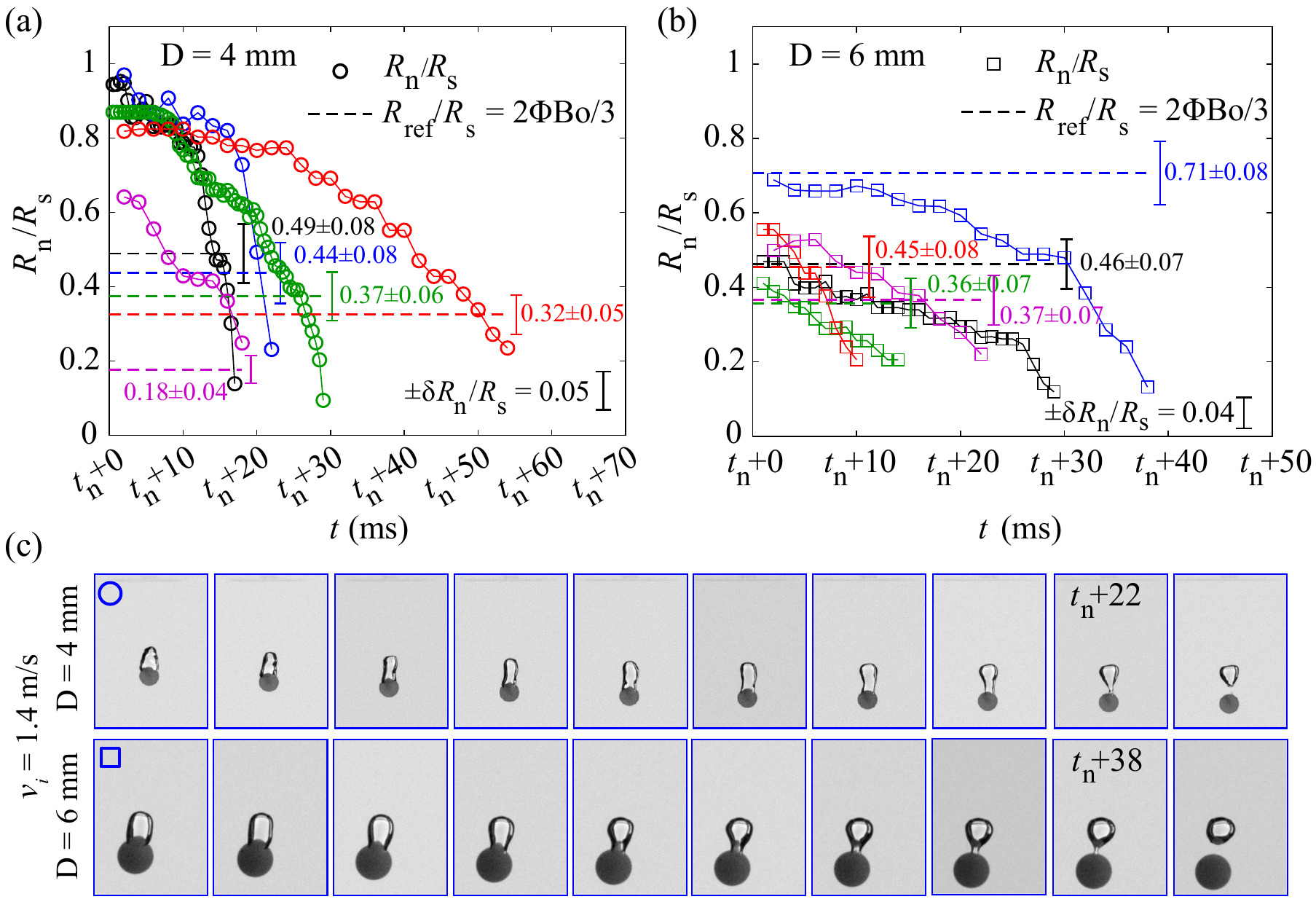}
    \caption{Neck evolution up to the secondary pinch-off for terminally descending spheres ($\rho^* = 1.12$, $\theta_a = 100^\circ$). Dimensionless neck radius $R_n(t)/R_s$ (symbols) and reference neck radius $R_\mathrm{ref}/R_\mathrm{s}=2\Phi\mathrm{Bo}/3$ (dashed lines), Eq.~(\ref{eq:iv_reference}), from the onset of the neck at $t_n$ until pinch-off, for (a) $D = 4$~mm and (b) $D = 6$~mm. Each color denotes an individual experiment; the dashed line of the same color is the reference radius of that experiment. Vertical bars at the end of each dashed line give the uncertainty of $R_{\mathrm{ref}}/R_s$ of that experiment, arising from the uncertainty of the displaced volume; the separate bar gives the edge-detection uncertainty of $R_n/R_s$ ($\pm1$~pixel per edge), which is the same for all points. (c) Image sequences of the neck thinning and the subsequent secondary pinch-off for $v_i = 1.4$~m/s, corresponding to the blue curves in (a) and (b). } 
\label{fig:neck_Descending}
\end{figure}

The measurements in Fig.~\ref{fig:neck_Descending} show that secondary pinch-off need not follow immediately from the neck width at formation. For the 4~mm spheres, the measured neck initially has $R_n/R_s\simeq0.65$--$0.97$ and subsequently contracts. The image sequences show changes in the three-phase contact-line position accompanying this contraction. These observations connect neck evolution with the changing attachment geometry but do not by themselves establish a constitutive relation between contact-line motion, static contact angle, and radial collapse.

\begin{figure}[ht!]
\centering
\includegraphics[width=\textwidth]{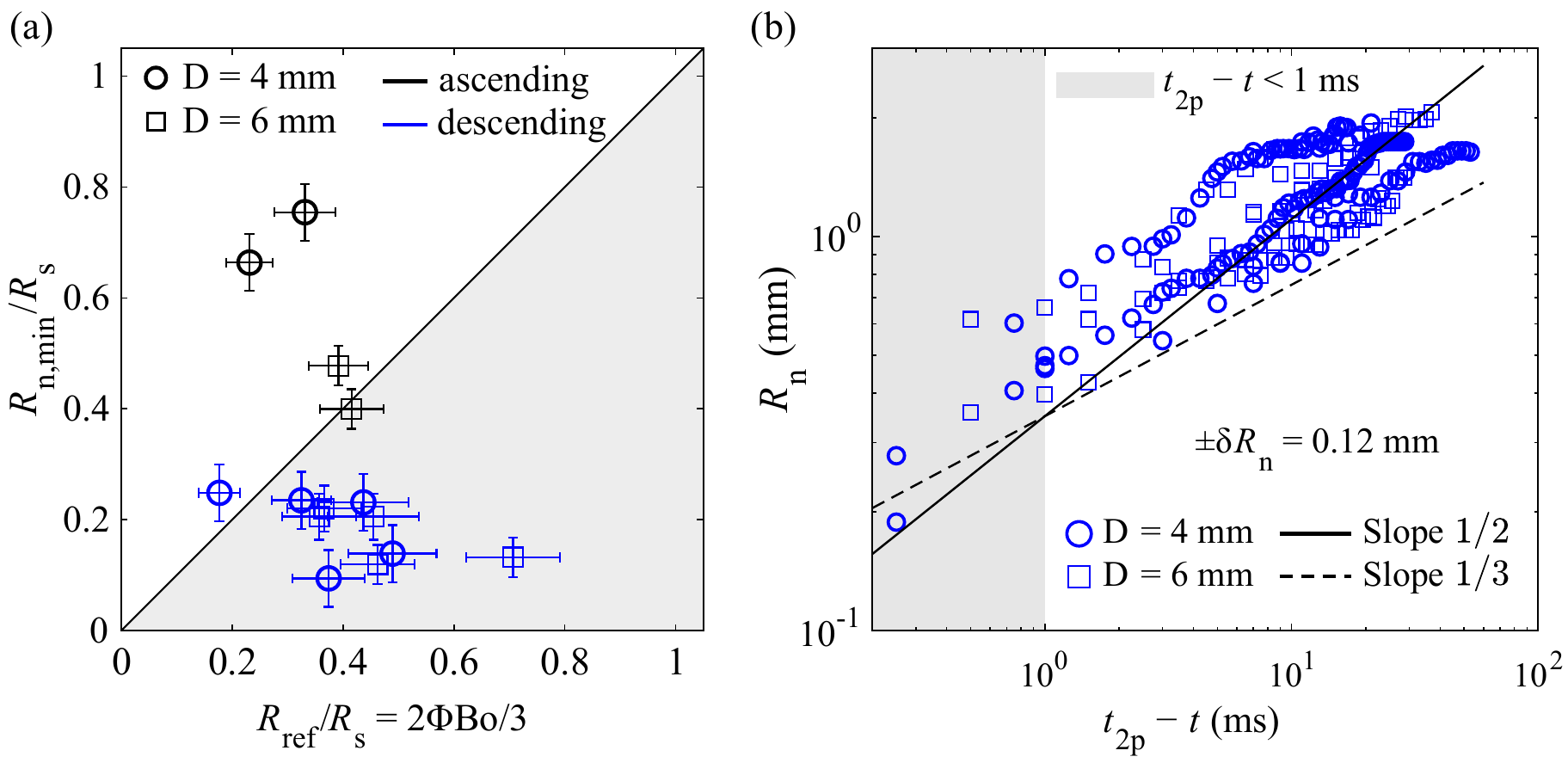}
    \caption{Neck radius of the air cavity prior to secondary pinch-off for spheres ($\rho^* = 1.12$, $\theta_a = 100^\circ$) of diameter $D = 4$~mm and $D = 6$~mm. (a) Minimum neck radius $R_{n,\min}/R_s$ reached in each experiment against the reference neck radius ratio $2\Phi\mathrm{Bo}/3$ of the same experiment, evaluated with its mean $V^*_{\mathrm{Disp}}$. Symbols represent individual experiments. For descending spheres, $R_{n,\min}$ is taken from the last frame before the secondary pinch-off. The solid line marks $R_{n,\min}/R_s = 2\Phi\mathrm{Bo}/3$; i.e., $R_{n,\min}<R_\mathrm{ref}$ in the shaded region below it. (b) Neck radius $R_n(t)$ as a function of the time remaining to the secondary pinch-off, $t_{2p} - t$. In (a), horizontal and vertical error bars give the uncertainties of $R_{\mathrm{ref}}/R_s$ (displaced volume) and $R_{n,\min}/R_s$ (edge detection), respectively. In (b), the shaded region marks $t_{2p} - t < 1$~ms, where the uncertainty of the pinch-off time, $\pm\Delta t/2$, $\pm0.25$--$1$~ms depending on the frame interval, exceeds 25\% of the remaining time; the value $\delta R_n$ gives the edge-detection uncertainty of the neck radius. } 
\label{fig:neck_collapse} 
\end{figure}

In Fig.~\ref{fig:neck_collapse}(a), the minimum resolved radius is compared with the reference radius from Eq.~\eqref{eq:iv_reference}. Retained and collapsing cavities generally occupy different sides of the equality $R_{n,\min}=R_{\mathrm{ref}}$. Within the propagated uncertainties (error bars), this holds for all cases except one ascending 6~mm case, which lies on the equality, and one descending 4~mm case, whose last resolved frame precedes pinch-off by up to 2~ms. For descending cases, however, the minimum is taken from the last resolved frame before pinch-off. A neck approaching zero radius will eventually cross any finite positive reference radius. This retrospective separation, therefore, does not establish a predictive threshold.

The reference values in Table~\ref{tab:secondary-comparison} illustrate the diameter dependence of the force comparison. With $\Phi\simeq0.57$, Eq.~\eqref{eq:iv_reference} gives $R_{\mathrm{ref}}/R_s\simeq0.21$, $0.47$ and $1.29$ for diameters of 4, 6 and 10~mm, respectively. These values can be compared with the illustrative neck ratio of approximately $0.45$. The proximity of the 6~mm value is suggestive, but it does not establish the cause of the observed variability. Likewise, $R_{\mathrm{ref}}/R_s>1$ does not prove that a neck cannot survive. The displayed trajectories in Fig.~\ref{fig:sphere 1.12} show ascent for the 4~mm cases, descent for the 10~mm cases, and both types of motion among the 6~mm cases. These observations remain experimental outcomes rather than predictions of Eq.~\eqref{eq:iv_reference}.
\begin{table*}[tbp]
\caption{Force comparison (Eq.~\eqref{eq:iv_reference}) versus observed terminal state.}
\label{tab:secondary-comparison}
\begin{ruledtabular}
\begin{tabular}{l c c c l }
Case (Fig.) & $D$ (mm) & Bo & $R_\mathrm{ref}/R_s$ & Observed  \\
\hline
SH, Fig.~\ref{fig:sphere 1.12} & 4  & 0.55 & 0.21 & Ascends (all)  \\
SH, Fig.~\ref{fig:sphere 1.12} & 6  & 1.22 & 0.47 & Both (stochastic)  \\
SH, Fig.~\ref{fig:sphere 1.12} & 10 & 3.45 & 1.31 ($>1$) & Descends (all)  \\
\hline
SH, $\mathrm{Fr}_i$ $=11.55$ & 6  & 1.22 & -- & Ascends, $V_{\mathrm{Disp}}^* = 1.42$  \\
SH, $\mathrm{Fr}_i = 8.94$  & 10 & 3.45 & -- & Descends, $V_{\mathrm{Disp}}^* = 1.02$  \\
\hline
Heavy ($\rho^*\ge3.26$), Fig.~\ref{fig:parameter_space}       & 4, 6, 10  & 0.55-3.45 &      & Never ascends  \\
Hydrophobic, Fig.~\ref{fig:neck_Descending} & 4, 6 & 0.55-1.22 & 0.18-0.71 & Descends       \\
\end{tabular}
\end{ruledtabular}
\end{table*}

Figure~\ref{fig:neck_collapse}(b) presents the neck radius against the time remaining to secondary pinch-off. The slopes $1/3$ and $1/2$ are reference guides. The scatter and temporal resolution do not support a unique collapse exponent or validation of the pressure terms in Eq.~\eqref{eq:iv_motion}. An approximately square-root dependence may be compatible with an inertial description, but does not identify the pressure closure or establish that the local cylindrical assumptions hold throughout the measured interval.

The retained volume provides a more direct connection to the subsequent sphere motion. In the illustrated 6~mm case at $\mathrm{Fr}_i=11.55$, $V_{\mathrm{Disp}}^*\simeq1.42$ exceeds $\rho^*=1.12$ and the sphere ascends. For the illustrated 10~mm case at $\mathrm{Fr}_i=8.94$, secondary pinch-off reduces the displaced volume to approximately $1.02$, below the density ratio, and the sphere continues to descend. These comparisons are consistent with the reduced buoyancy condition in Eq.~(\ref{eq:condition for ascending}). They do not imply that retained cavities of sufficient volume are impossible at larger density ratios; no such ascending cases were observed within the investigated range for $\rho^*\geq3.26$.

The measurements therefore support a sequence in which cavity and contact-line evolution determine how much air remains attached, and the retained volume changes the buoyancy available to reverse the sphere motion. Predicting that sequence requires the neck pressure and geometry to be coupled to the evolving cavity and sphere trajectory. The force comparison provides a way to organize the present observations, while the local radial balance identifies the unresolved quantities needed to predict secondary pinch-off.

\section{Conclusions}
\label{sec:conclusion}

The water entry of hydrophobic and superhydrophobic spheres was investigated using high-speed imaging to measure sphere trajectories, cavity geometry, and retained displaced volume. Within the investigated parameter range, secondary pinch-off was followed by continued descent; no sphere was observed to return to the surface after this event. When sufficient air remained attached, the combined displacement of the sphere and cavity produced buoyancy exceeding the sphere weight, allowing the downward motion to stop and reverse. The condition \(V_{\mathrm{Disp}}^*>\rho^*\) therefore provides the buoyancy requirement for reversal within the reduced force description.

Surface wettability influenced the persistence and evolution of the attached cavity. The image sequences show neck thinning accompanied by changes in the three-phase contact-line position. For the illustrated lightest superhydrophobic spheres, the 4 mm spheres ascended, the 10 mm spheres descended, and repeated 6 mm experiments exhibited both outcomes. These observations demonstrate that the trajectory depends on the air retained after cavity evolution, rather than on wettability alone.

The neck measurements further show that its width at formation does not uniquely determine whether secondary pinch-off occurs. Some cavities initially had a broad neck that subsequently contracted before rupture. A simplified force comparison defines a reference neck radius ratio against which this evolution can be assessed. The minimum resolved neck-radius ratios of retained and collapsing cavities generally fall on opposite sides of the corresponding reference-radius ratio. This agreement supports its use as a phenomenological interpretation but does not establish a predictive stability criterion. The principal experimental result is the connection between secondary neck evolution, loss of retained air, and the subsequent direction of sphere motion.

\begin{acknowledgments}
CT acknowledges support from the Indian Institute of Technology Madras through the appointment as the Henry Ford Chair Professorship. PSM acknowledges the V. Ganesan Faculty Fellowship received from the Indian Institute of Technology Madras.
\end{acknowledgments}

\section*{Declaration of interests}
The authors report no conflicts of interest.

\nocite{apsrev42Control}
\bibliographystyle{apsrev4-2}
\bibliography{references.bib}

@CONTROL{apsrev42Control,author="08",editor="1",pages="0",title="0",year="1"}

@article{gekle2009approach,
  title={Approach to universality in axisymmetric bubble pinch-off},
  author={Gekle, Stephan and Snoeijer, Jacco H and Lohse, Detlef and van der Meer, Devaraj},
  journal={Physical Review E},
  volume={80},
  number={3},
  pages={036305},
  year={2009},
  publisher={APS}
}

@article{eggers2007theory,
  title={Theory of the collapsing axisymmetric cavity},
  author={Eggers, J and Fontelos, MA and Leppinen, D and Snoeijer, JH},
  journal={Physical Review Letters},
  volume={98},
  number={9},
  pages={094502},
  year={2007},
  publisher={APS}
}

@phdthesis{tan2016experimental,
  title={An experimental study on the entry of spheres into a two-layer system of immiscible liquids},
  author={Tan, Benedict Chuan-Wei},
  year={2016},
  school={University of Warwick}
}

@article{Maru1971BehaviorInterface,
    title = {{Behavior of a rigid sphere at a liquid-liquid interface}},
    year = {1971},
    journal = {Chemical Engineering Science},
    author = {Maru, H. C. and Wasan, D. T. and Kintner, R. C.},
    number = {10},
    pages = {1615--1628},
    volume = {26}
}

@article{Wang2015BehaviorDropletsurface,
    title = {{Behavior of hydrophobic micron particles impacting on dropletsurface}},
    year = {2015},
    journal = {Atmospheric Environment},
    author = {Wang, Ao and Song, Qiang and Yao, Qiang},
    pages = {1--8},
    volume = {115},
    publisher = {Elsevier Ltd}
}

@article{Marston2011BubbleSurfaces,
    title = {{Bubble entrapment during sphere impact onto quiescent liquid surfaces}},
    year = {2011},
    journal = {Journal of Fluid Mechanics},
    author = {Marston, J. O. and Vakarelski, I. U. and Thoroddsen, S. T.},
    number = {May 1951},
    pages = {660--670},
    volume = {680}
}

@article{Liu2007BuoyantWater,
    title = {{Buoyant force and sinking conditions of a hydrophobic thin rod floating on water}},
    year = {2007},
    journal = {Physical Review E - Statistical, Nonlinear, and Soft Matter Physics},
    author = {Liu, Jian Lin and Feng, Xi Qiao and Wang, Gang Feng},
    number = {6},
    pages = {1--9},
    volume = {76}
}

@phdthesis{truscott2009cavity,
       author = {Truscott, Tadd Trevor},
       title = {Cavity dynamics of water entry for spheres and ballistic projectiles},
       school = {Massachusetts Institute of Technology},
       year = {2009},
       address = {Cambridge, MA},
       type = {Ph.{D}. thesis},
       month = {June}
}

@article{Truscott2006CavitySurface,
    title = {{Cavity formation in the wake of a spinning sphere impacting the free surface}},
    year = {2006},
    journal = {Physics of Fluids},
    author = {Truscott, Tadd T. and Techet, Alexandra H.},
    number = {9},
    pages = {22--23},
    volume = {18}
}

@article{Viswanadam2012ContactSurfaces,
    title = {{Contact angles of drops on curved superhydrophobic surfaces}},
    year = {2012},
    journal = {Journal of Colloid and Interface Science},
    author = {Viswanadam, Goutham and Chase, George G.},
    number = {1},
    pages = {472--477},
    volume = {367},
    publisher = {Elsevier Inc.},
    url = {http://dx.doi.org/10.1016/j.jcis.2011.11.004},
    doi = {10.1016/j.jcis.2011.11.004},
    issn = {00219797}
}

@inproceedings{magos2021ContactSurfaces,
  title={Contact angles on spherical hydrophilic surfaces},
  author={Magos, Istv{\'a}n and B{\u{a}}lan, Corneliu},
  booktitle={2021 12th International Symposium on Advanced Topics in Electrical Engineering (ATEE)},
  pages={1--4},
  year={2021},
  organization={IEEE}
}

@article{vakarelski2011drag,
  title={Drag reduction by {L}eidenfrost vapor layers},
  author={Vakarelski, Ivan U and Marston, Jeremy O and Chan, Derek YC and Thoroddsen, Sigurdur T},
  journal={Physical Review Letters},
  volume={106},
  number={21},
  pages={214501},
  year={2011},
  publisher={APS}
}

@article{Duclaux2007DynamicsCavities,
    title = {{Dynamics of transient cavities}},
    year = {2007},
    journal = {Journal of Fluid Mechanics},
    author = {Duclaux, V. and Caill{\'{e}}, F. and Duez, C. and Ybert, C. and Bocquet, L. and Clanet, C.},
    pages = {1--19},
    volume = {591}
}

@article{Chen2018EntrappingInterface,
    title = {{Entrapping an impacting particle at a liquid-gas interface}},
    year = {2018},
    journal = {Journal of Fluid Mechanics},
    author = {Chen, Han and Liu, Hao Ran and Lu, Xi Yun and Ding, Hang},
    pages = {1073--1084},
    volume = {841}
}

@phdthesis{laverty2005experimental,
  title={Experimental hydrodynamics of spherical projectiles impacting on a free surface using high speed imaging techniques},
  author={Laverty, Stephen Michael},
  year={2005},
  school={Massachusetts Institute of Technology}
}

@article{Guleria2021ExperimentalSphere,
    title = {{Experimental insights on the water entry of hydrophobic sphere}},
    year = {2021},
    journal = {Physics of Fluids},
    author = {Guleria, Sharey Deep and Dhar, Atul and Patil, Dhiraj V.},
    number = {10},
    volume = {33},
    publisher = {AIP Publishing LLC}
}

@article{li2019experimental,
  title={Experimental study on water entry of spheres with different surface wettability},
  author={Li, Daqin and Zhang, Jiayue and Zhang, Mindi and Huang, Biao and Ma, Xiaojian and Wang, Guoyu},
  journal={Ocean Engineering},
  volume={187},
  pages={106123},
  year={2019},
  publisher={Elsevier}
}

@article{Mehri2020HydrodynamicEntry,
    title = {{Hydrodynamic characteristics of heated/non-heated and grooved/un-grooved spheres during free-surface water entry}},
    year = {2020},
    journal = {Journal of Fluids and Structures},
    author = {Mehri, Ali and Akbarzadeh, Pooria},
    pages = {103100},
    volume = {97},
    publisher = {Elsevier Inc.}
}

@article{Lee2008ImpactWater,
    title = {{Impact of a superhydrophobic sphere onto water}},
    year = {2008},
    journal = {Langmuir},
    author = {Lee, Duck Gyu and Kim, Ho Young},
    number = {1},
    pages = {142--145},
    volume = {24}
}

@article{Extrand2012IndirectSurfaces,
    title = {{Indirect methods to measure wetting and contact angles on spherical convex and concave surfaces}},
    year = {2012},
    journal = {Langmuir},
    author = {Extrand, C. W. and Moon, Sung In},
    number = {20},
    pages = {7775--7779},
    volume = {28}
}

@article{Gilbarg1948InfluenceWater,
    title = {{Influence of atmospheric pressure on the phenomena accompanying the entry of spheres into water}},
    year = {1948},
    journal = {Journal of Applied Physics},
    author = {Gilbarg, David and Anderson, Robert A.},
    number = {2},
    pages = {127--139},
    volume = {19}
}

@article{Vakarelski2014LeidenfrostWater,
    title = {{Leidenfrost vapour layer moderation of the drag crisis and trajectories of superhydrophobic and hydrophilic spheres falling in water}},
    year = {2014},
    journal = {Soft Matter},
    author = {Vakarelski, Ivan U. and Chan, Derek Y.C. and Thoroddsen, Sigurdur T.},
    number = {31},
    pages = {5662--5668},
    volume = {10}
}

@article{Vakarelski2019Mobile-surfaceStronger,
    title = {{Mobile-surface bubbles and droplets coalesce faster but bounce stronger}},
    year = {2019},
    journal = {Science Advances},
    author = {Vakarelski, Ivan U. and Yang, Fan and Tian, Yuan Si and Li, Er Qiang and Chan, Derek Y.C. and Thoroddsen, Sigurdur T.},
    number = {10},
    pages = {1--8},
    volume = {5}
}

@article{Galeano-Rios2017Non-wettingDroplets,
    title = {{Non-wetting impact of a sphere onto a bath and its application to bouncing droplets}},
    year = {2017},
    journal = {Journal of Fluid Mechanics},
    author = {Galeano-Rios, Carlos A. and Milewski, Paul A. and Vanden-Broeck, J. M.},
    pages = {97--127},
    volume = {826}
}

@article{Vakarelski2017Self-determinedCavities,
    title = {{Self-determined shapes and velocities of giant near-zero drag gas cavities}},
    year = {2017},
    journal = {Science Advances},
    author = {Vakarelski, Ivan U. and Klaseboer, Evert and Jetly, Aditya and Mansoor, Mohammad M. and Aguirre-Pablo, Andres A. and Chan, Derek Y.C. and Thoroddsen, Sigurdur T.},
    number = {9},
    pages = {1--8},
    volume = {3},
    pmid = {28913434}
}

@article{Mansoor2017Stable-streamlinedSpheres,
    title = {{Stable-streamlined and helical cavities following the impact of Leidenfrost spheres}},
    year = {2017},
    journal = {Journal of Fluid Mechanics},
    author = {Mansoor, M. M. and Vakarelski, I. U. and Marston, J. O. and Truscott, T. T. and Thoroddsen, S. T.},
    pages = {716--754},
    volume = {823}
}

@article{Vakarelski2019Stable-streamlinedWater,
    title = {{Stable-streamlined cavities following the impact of non-superhydrophobic spheres on water}},
    year = {2019},
    journal = {Soft Matter},
    author = {Vakarelski, Ivan U. and Jetly, Aditya and Thoroddsen, Sigurdur T.},
    number = {31},
    pages = {6278--6287},
    volume = {15},
    publisher = {Royal Society of Chemistry}
}

@article{Vella2007SurfaceImpact,
    title = {{Surface tension dominated impact}},
    year = {2007},
    journal = {Physics of Fluids},
    author = {Vella, Dominic and Metcalfe, Paul D.},
    number = {7},
    volume = {19}
}

@article{Vella2006TheObjects,
    title = {{The load supported by small floating objects}},
    year = {2006},
    journal = {Langmuir},
    author = {Vella, Dominic and Lee, Duck Gyu and Kim, Ho Young},
    number = {14},
    pages = {5979--5981},
    volume = {22}
}

@article{Truscott2012UnsteadyEntry,
    title = {{Unsteady forces on spheres during free-surface water entry}},
    year = {2012},
    journal = {Journal of Fluid Mechanics},
    author = {Truscott, Tadd T. and Epps, Brenden P. and Techet, Alexandra H.},
    pages = {173--210},
    volume = {704}
}

@article{May1952VerticalWater,
    title = {{Vertical entry of missiles into water}},
    year = {1952},
    journal = {Journal of Applied Physics},
    author = {May, Albert},
    number = {12},
    pages = {1362--1372},
    volume = {23},
}

@article{Extrand2002WaterSurfaces,
    title = {{Water contact angles and hysteresis of polyamide surfaces}},
    year = {2002},
    journal = {Journal of Colloid and Interface Science},
    author = {Extrand, C. W.},
    number = {1},
    pages = {136--142},
    volume = {248},
    pmid = {16290514}
}

@article{Mehri2021WaterVelocity,
    title = {{Water entry of grooved spheres: Effect of the number of grooves and impact velocity}},
    year = {2021},
    journal = {Journal of Fluids and Structures},
    author = {Mehri, Ali and Akbarzadeh, Pooria},
    pages = {103198},
    volume = {100},
    publisher = {Elsevier Inc.}
}

@article{Truscott2014WaterProjectiles,
    title = {{Water entry of projectiles}},
    year = {2014},
    journal = {Annual Review of Fluid Mechanics},
    author = {Truscott, Tadd T. and Epps, Brenden P. and Belden, Jesse},
    pages = {355--378},
    volume = {46}
}

@article{Aristoff2009WaterSpheres,
    title = {{Water entry of small hydrophobic spheres}},
    year = {2009},
    journal = {Journal of Fluid Mechanics},
    author = {Aristoff, Jeffrey M. and Bush, John W.M.},
    pages = {45--78},
    volume = {619},
}

@article{Speirs2019WaterAnglesb,
    title = {{Water entry of spheres with various contact angles}},
    year = {2019},
    journal = {Journal of Fluid Mechanics},
    author = {Speirs, Nathan B. and Mansoor, Mohammad M. and Belden, Jesse and Truscott, Tadd T.},
    pages = {1--13},
    volume = {862}
}

@article{techet2011water,
  title={Water entry of spinning hydrophobic and hydrophilic spheres},
  author={Techet, AH and Truscott, TT},
  journal={Journal of Fluids and Structures},
  volume={27},
  number={5-6},
  pages={716--726},
  year={2011},
  publisher={Elsevier}
}

@article{Mansoor2014WaterFormation,
    title = {{Water entry without surface seal: Extended cavity formation}},
    year = {2014},
    journal = {Journal of Fluid Mechanics},
    author = {Mansoor, M. M. and Marston, J. O. and Vakarelski, I. U. and Thoroddsen, S. T.},
    pages = {295--326},
    volume = {743}
}

@inproceedings{epps2010evaluating,
  title={Evaluating derivatives of experimental data using smoothing splines},
  author={Epps, Brenden P and Truscott, Tadd T and Techet, Alexandra H},
  booktitle={Proceedings of Mathematical Methods in Engineering International Symposium. MMEI, Lisbon Portugal},
  pages={29--38},
  year={2010}
}

@article{worthington1883impact,
  title={On impact with a liquid surface},
  author={Worthington, Arthur Mason},
  journal={Proceedings of the Royal Society of London},
  volume={34},
  number={220-223},
  pages={217--230},
  year={1883},
  publisher={The Royal Society London}
}

@article{weisensee2016water,
  title={Water droplet impact on elastic superhydrophobic surfaces},
  author={Weisensee, Patricia B and Tian, Junjiao and Miljkovic, Nenad and King, William P},
  journal={Scientific Reports},
  volume={6},
  number={1},
  pages={30328},
  year={2016},
  publisher={Nature Publishing Group UK London}
}

@article{richardson1948impact,
  title={The impact of a solid on a liquid surface},
  author={Richardson, EG},
  journal={Proceedings of the Physical Society},
  volume={61},
  number={4},
  pages={352},
  year={1948},
  publisher={IOP Publishing}
}

@book{may1965study,
  title={A study of the water-entry cavity},
  author={May, Albert and Hoover, William R},
  year={1965},
  publisher={United States Naval Ordance Laboratory}
}

@inproceedings{asfar1987rigid,
  title={Rigid-body water impact at shallow angles of incidence},
  author={Asfar, K and Moore, S},
  booktitle={Proceedings of the Sixth International Offshore Mechanics and Arctic Engineering Symposium},
  pages={105--112},
  year={1987}
}

@article{aristoff2008water,
  title={The water-entry cavity formed by low {B}ond number impacts},
  author={Aristoff, Jeffrey M and Truscott, Tadd T and Techet, Alexandra H and Bush, John WM},
  journal={Physics of Fluids},
  volume={20},
  number={9},
  year={2008},
  publisher={AIP Publishing}
}

@article{vella2015floating,
  title={Floating versus sinking},
  author={Vella, Dominic},
  journal={Annual Review of Fluid Mechanics},
  volume={47},
  number={1},
  pages={115--135},
  year={2015},
  publisher={Annual Reviews}
}

@article{Billa_Josyula_Tropea_Mahapatra_2025, title={Motion of a rigid sphere penetrating a deep pool}, volume={1012}, journal={Journal of Fluid Mechanics}, author={Billa, Prasanna Kumar and Josyula, Tejaswi and Tropea, Cameron and Mahapatra, Pallab Sinha}, year={2025}, pages={A5}}

@article{billa2026entry,
  title={Entry and penetration of a superhydrophobic sphere into a deep pool},
  author={Billa, Prasanna Kumar and Tropea, Cameron and Mahapatra, Pallab Sinha},
  journal={Physical Review Fluids},
  volume={11},
  number={1},
  pages={014304},
  year={2026},
  publisher={APS}
}

@article{yuh2000design,
  title={Design and control of autonomous underwater robots: A survey},
  author={Yuh, Junku},
  journal={Autonomous Robots},
  volume={8},
  number={1},
  pages={7--24},
  year={2000},
  publisher={Springer}
}

@article{neira2021review,
  title={Review on unmanned underwater robotics, structure designs, materials, sensors, actuators, and navigation control},
  author={Neira, Javier and Sequeiros, Cristhel and Huamani, Richard and Machaca, Elfer and Fonseca, Paola and Nina, Wilder},
  journal={Journal of Robotics},
  volume={2021},
  number={1},
  pages={5542920},
  year={2021},
  publisher={Wiley Online Library}
}

@article{jung2025water,
  title={Water entry and exit in nature},
  author={Jung, Sunghwan},
  journal={Interface Focus},
  volume={15},
  number={2},
  pages={20240055},
  year={2025},
  publisher={The Royal Society}
}

@article{seddon2006review,
  title={Review of water entry with applications to aerospace structures},
  author={Seddon, CM and Moatamedi, M},
  journal={International Journal of Impact Engineering},
  volume={32},
  number={7},
  pages={1045--1067},
  year={2006},
  publisher={Elsevier}
}

@article{zhao2025bioinspired,
  title={Bioinspired hydrophilic strip patterning on superhydrophobic spheres for stabile underwater cavities and enhanced drag reduction},
  author={Zhao, J and He, PC and Yang, JH and Liu, YJ and Li, S and Qu, JP and Cheng, ZJ and Mo, JL},
  journal={Applied Materials Today},
  volume={44},
  pages={102770},
  year={2025},
  publisher={Elsevier}
}

@article{zhu2022sustaining,
  title={Sustaining robust cavities with slippery liquid--liquid interfaces},
  author={Zhu, Suwan and Wu, Tao and Bian, Yucheng and Chen, Chao and Zhang, Yiyuan and Li, Jiawen and Wu, Dong and Hu, Yanlei and Chu, Jiaru and Li, Erqiang and others},
  journal={Advanced Science},
  volume={9},
  number={7},
  pages={2103568},
  year={2022},
  publisher={Wiley Online Library}
}

\end{document}